\documentclass[12pt]{article}
\pdfoutput=1

\usepackage{epsfig}
\usepackage{psfrag}
\usepackage{latexsym}
\usepackage{indentfirst}
\usepackage{fancyhdr}
\usepackage{dsfont}
\usepackage{amssymb}
\usepackage{amsmath}
\usepackage{amsfonts}
\usepackage{pifont}
\usepackage{cite}
\usepackage{bbold}
\usepackage{color}
\usepackage{colordvi}
\usepackage{fancybox}
\usepackage[footnotesize]{caption2}
\usepackage{graphicx}
\usepackage[center,footnotesize,hang]{subfigure}
\usepackage{bbm}
\usepackage{url}
\usepackage{multirow}
\usepackage{array}
\usepackage{arydshln}
\usepackage{ulem}
\usepackage{enumitem}
\newcommand{\PreserveBackslash}[1]{\let\temp=\\#1\let\\=\temp}
\newcolumntype{C}[1]{>{\PreserveBackslash\centering}p{#1}}
\newcolumntype{R}[1]{>{\PreserveBackslash\raggedleft}p{#1}}
\newcolumntype{L}[1]{>{\PreserveBackslash\raggedright}p{#1}}
\allowdisplaybreaks  % automated pagination

\newcommand{\bq}{\begin{eqnarray}}
\newcommand{\nq}{\end{eqnarray}}

\allowdisplaybreaks[1]

\def\simgt{\mathrel{\lower2.5pt\vbox{\lineskip=0pt\baselineskip=0pt
           \hbox{$>$}\hbox{$\sim$}}}}
\def\simlt{\mathrel{\lower2.5pt\vbox{\lineskip=0pt\baselineskip=0pt
           \hbox{$<$}\hbox{$\sim$}}}}

\newcommand{\be}{\begin{eqnarray}}
\newcommand{\ee}{\end{eqnarray}}

\def\bea{\begin{eqnarray}}
\def\eea{\end{eqnarray}}

\begin{document}

\title{\hfill ~\\[0mm]
        \textbf{Lepton Flavor Mixing and CP Symmetry}}

\date{}

\author{\\[1mm]Peng Chen\footnote{E-mail: {\tt pche@mail.ustc.edu.cn}}~,~~Cai-Chang Li\footnote{E-mail: {\tt lcc0915@mail.ustc.edu.cn}}~,~~Gui-Jun Ding\footnote{E-mail: {\tt dinggj@ustc.edu.cn}}\\ \\
\it{\small Department of Modern Physics, University of Science and
    Technology of China,}\\
  \it{\small Hefei, Anhui 230026, China}\\[4mm] }
\maketitle

\begin{abstract}
\noindent

The strategy of constraining the lepton flavor mixing from remnant CP symmetry is investigated in a rather general way. The neutrino mass matrix generally admits four remnant CP transformations which can be derived from the measured lepton mixing matrix in the charged lepton diagonal basis. Conversely, the lepton mixing matrix can be reconstructed from the postulated remnant CP transformations. All mixing angles and CP violating phases can be completely determined by the full set of remnant CP transformations or three of them. When one or two remnant CP transformations are preserved, the resulting lepton mixing matrix would depend on three real parameters or one real parameter respectively in addition to the parameters characterizing the remnant CP, and the concrete form of the mixing matrix is presented. The phenomenological predictions for the mixing parameters are discussed. The conditions leading to vanishing or maximal Dirac CP violation are studied.

\end{abstract}
\thispagestyle{empty}
\vfill

\newpage
\setcounter{page}{1}

%%%%%%%%%%%%%%%%%%%%%% Introduction %%%%%%%%%%%%%%%%%%%%%%%

\section{\label{sec:introduction}Introduction}
\setcounter{equation}{0}

The origin of flavor mixing is one of longstanding open questions in particle physics. Firstly motivated by the well-known tri-bimaximal mixing~\cite{Harrison:2002er}, a considerable effort has been devoted to understanding lepton mixing from a discrete flavor symmetry which is spontaneously broken down to two different residual subgroups in the neutrino and the charged lepton sectors. Please see Refs.~\cite{Altarelli:2010gt,Ishimori:2010zr,Grimus:2012dk,King:2013eh,King:2014nza} for review of discrete flavor symmetries and their application in model building aspects. So far a complete classification of all possible lepton mixing which could be derived from a finite flavor symmetry group under the hypothesis of Majorana neutrino has been accomplished~\cite{Fonseca:2014koa}.
Among the complete list of mixing patterns achievable, only the Pontecorvo-Maki-Nakagawa-Sakata (PMNS) matrix with the second column being $(1,1,1)^{T}/\sqrt{3}$ can be compatible with experimental data, and the Dirac CP-violating phase is fixed to trivial. Moreover, the Majorana phases are indeterminate as the neutrino masses are unconstrained by flavor symmetry.

On the experimental side, the precise measurements of the reactor mixing angle $\theta_{13}$ by T2K~\cite{Abe:2011sj}, MINOS~\cite{Adamson:2011qu}, DOUBLE-CHOOZ~\cite{Abe:2011fz}, RENO~\cite{Ahn:2012nd} and DAYA-BAY~\cite{An:2012eh} reactor neutrino experiments is one of the most significant discoveries in recent years. The sizable $\theta_{13}\sim9^{\circ}$ opens the gateway to access two remaining unknown parameters in the neutrino sector: the neutrino mass hierarchy and the leptonic Dirac CP phase $\delta_{CP}$. If neutrinos are Majorana particles, there are two additional Majorana CP phases which can play a critical role in the neutrinoless double beta decay, and we know nothing about their values so far. The T2K collaboration reported a weak evidence for nonzero $\delta_{CP}\sim3\pi/2$~\cite{Abe:2013hdq}, and some indications of nontrivial $\delta_{CP}$ are starting to appear in global analysis of neutrino oscillation data~\cite{Capozzi:2013csa,Forero:2014bxa,Gonzalez-Garcia:2014bfa}. Needless to say, probing CP violation in the lepton sector would help deepen our understanding of the universe. Some long-baseline neutrino oscillation experiments such as LBNE~\cite{Adams:2013qkq}, LBNO~\cite{::2013kaa} and Hyper-Kamiokande~\cite{Abe:2011ts} have been proposed to precisely measure the lepton mixing parameters in particular the Dirac CP phase $\delta_{CP}$.

If the signal of CP violation is observed in future neutrino oscillation apparatuses, the paradigm of the flavor symmetry would be disfavored. Moreover, in light of the hints for maximal Dirac CP violation $\delta_{CP}\sim3\pi/2$, it is imperative and significative to be able to understand the observed lepton mixing angles and meanwhile predict the values of CP phases from certain underlying principles. It is notable that CP symmetry was found to impose strong constraints on the fermion mass matrices nearly thirty years ago~\cite{Ecker:1981wv,Grimus:1995zi}. A typical simple CP transformation is the so-called $\mu-\tau$ reflection
under which $\nu_{\mu}$ and $\nu_{\tau}$ transform into the CP-conjugate of each other~\cite{Harrison:2002kp,Grimus:2003yn,Farzan:2006vj}. A neutrino mass matrix fulfilling the $\mu-\tau$ reflection symmetry immediately gives rise to both maximal atmospheric mixing angle $\theta_{23}$ and maximal CP violation $\cos\delta_{CP}=0$. In recent years, it is found that the $\mu-\tau$ reflection can naturally appear when CP symmetry is imposed together with the widely studied $S_4$ flavor symmetry~\cite{Mohapatra:2012tb,Feruglio:2012cw,Ding:2013hpa}. Furthermore, phenomenological models in which the desired breaking patterns of the flavor and CP symmetries are achieved dynamically have been constructed~\cite{Ding:2013hpa,Feruglio:2013hia,Luhn:2013lkn,Li:2013jya,Li:2014eia}. The interplay between CP symmetry with the flavor symmetries $A_4$~\cite{Ding:2013bpa} and $T^{\prime}$~\cite{Chen:2009gf,Girardi:2013sza} has been investigated as well. When CP symmetry is combined with $\Delta(48)$~\cite{Ding:2013nsa} or $\Delta(96)$~\cite{Ding:2014ssa} flavor symmetry, CP transformations distinct from $\mu-\tau$ reflection can be produced such that $\delta_{CP}$ can be non-maximal. It turns out that both mixing angles and CP phases depend on only one common free parameter in that case. Recently the possible lepton mixing patterns derived from CP symmetry and the $\Delta(6n^2)$ or $\Delta(3n^2)$ flavor symmetry group series have been analyzed~\cite{King:2014rwa,Hagedorn:2014wha,Ding:2014ora}, the experimentally preferred values of the mixing angles can be accommodated very well, and the corresponding phenomenological implications in neutrinoless double decay are discussed~\cite{Ding:2014ora}. Note that it is highly nontrivial to consistently define the CP symmetry in the context of a finite flavor symmetry~\cite{Holthausen:2012dk,Chen:2014tpa}. There are more than one theoretical approachs dealing with flavor symmetry and CP violation~\cite{Branco:1983tn}.

In this work, we shall only concentrate on CP symmetry, and show that the lepton mixing matrix can be reconstructed from the remnant CP symmetry. We will derive the explicit form of the PMNS matrix when one or two residual CP transformations are preserved. Phenomenological implications for the lepton flavor mixing parameters are discussed in detail. Compared with flavor symmetry paradigm, both mixing angles and CP phases can be predicted by remnant CP, and the observed value of the reactor mixing angles can be easily accommodated.

The paper is organized as follows. In section~\ref{sec:remnant_symmetry}, remnant flavor symmetry and remnant CP transformations of the neutrino mass matrix are analyzed in the charged lepton diagonal basis. In section~\ref{sec:reconstruction_PMNS}, we show that the lepton mixing matrix can be reconstructed from the presumed remnant CP transformations. If two (or one) remnant CP transformations are preserved, the explicit form of the PMNS matrix is derived, and it depends on one (or three) free real parameters in addition to the parameters of the remnant CP. In section~\ref{sec:phenomenology}, the phenomenological predictions for the lepton mixing parameters are discussed, and we search for conditions of zero or maximal Dirac CP violation. Finally we summarize our results in section~\ref{sec:conclusion}.

\section{\label{sec:remnant_symmetry}Remnant symmetries of the mass matrices}
\setcounter{equation}{0}

In this section, we shall clarify the remnant flavor symmetry and remnant CP symmetry of the lepton mass matrices. We shall assume throughout this paper that the neutrinos are Majorana particles. The lepton mass terms obtained after symmetry breaking are of the following form:
\begin{equation}
\label{eq:mass_Lagrange}\mathcal{L}_{mass}=-\overline{l}_{R}m_{l}l_{L}+\frac{1}{2}\nu^{T}_{L}C^{-1}m_{\nu}\nu_{L}+h.c.\,,
\end{equation}
where $C$ is the charge-conjugation matrix, $l_L \equiv (e_L,\mu_L,\tau_L)^T$ and $l_R \equiv (e_R,\mu_R,\tau_R)^T$ denote the three generation left and right-handed charged lepton fields respectively, and $\nu_L \equiv (\nu_{eL}, \nu_{\mu L}, \nu_{\tau L})^T$ is the three left-handed neutrino fields. The Majorana neutrino mass matrix $m_{\nu}$ is symmetric. Since the mixing matrix only relates to left-handed fermions in standard model, as usual we construct the hermitian mass matrix $\mathcal{M}_{l}\equiv m^{\dagger}_{l}m_{l}$ which connects left-handed charged leptons on both sides. We denote the unitary diagonalization matrix of $\mathcal{M}_{l}$ and $m_{\nu}$ by $U_{l}$ and $U_{\nu}$ respectively, i.e.,
\begin{equation}
U^{\dagger}_{l}\mathcal{M}_{l}U_{l}=\text{diag}\left(m^2_{e}, m^2_{\mu}, m^2_{\tau}\right),\qquad U^{T}_{\nu}m_{\nu}U_{\nu}=\text{diag}\left(m_1, m_2, m_3\right)\equiv m_{diag}\,,
\end{equation}
where the light neutrino masses $m_{i} (i=1,2,3)$ are real and non-negative. The lepton mixing matrix is the mismatch between neutrino and charged lepton diagonalization matrices,
\begin{equation}
U_{PMNS}=U^{\dagger}_{l}U_{\nu}\,.
\end{equation}
Without loss of generality we shall choose to work in the basis where $\mathcal{M}_{l}$ is diagonal. Then $U_{l}$ would reduce to a unit matrix and the lepton mixing completely comes from the neutrino sector with $U_{PMNS}=U_{\nu}$. The general form of $m_{\nu}$ is
\begin{equation}
\label{eq:mnu}m_{\nu}=U^{*}_{PMNS}\;m_{diag}\;U^{\dagger}_{PMNS}\,.
\end{equation}
Firstly let's determine the remnant flavor symmetries $G_\nu$ and $G_l$ of the neutrino and charged lepton mass terms. A unitary transformation $\nu_{L}\rightarrow G_{\nu}\nu_{L}$ of the left-handed Majorana neutrino leads to the transformation of the neutrino mass matrix $m_{\nu}\rightarrow G^{T}_{\nu}m_{\nu}G_{\nu}$. $G_{\nu}$ is a flavor symmetry if and only if $m_{\nu}$ is invariant, i.e.,
\begin{equation}
G^{T}_{\nu}m_{\nu}G_{\nu}=m_{\nu}
\end{equation}
Substituting the expression of $m_{\nu}$ in Eq.~\eqref{eq:mnu} into this invariant condition and considering that the three light neutrino masses $m_{i}$ are non-degenerate~\footnote{Here we assume that the three light neutrino masses are non-vanishing. If the lightest neutrino is massless, then one diagonal entry ``$\pm1$'' could be replaced with an arbitrary phase factor in Eq.~\eqref{eq:remnant_Fla_cons}.}, we obtain
\begin{equation}
\label{eq:remnant_Fla_cons}U^{\dagger}_{PMNS}G_{\nu}U_{PMNS}=\text{diag}\left(\pm1, \pm1, \pm1\right)\,.
\end{equation}
As an overall $-1$ factor of $G_{\nu}$ is irrelevant, there are essentially four solutions for $G_{\nu}$,
\begin{equation}
G_{i}=U_{PMNS}\;d_{i}U^{\dagger}_{PMNS},\qquad i=1,2,3,4\,,
\end{equation}
where
\begin{eqnarray}
\nonumber&&d_1=\text{diag}\left(1,-1,-1\right),\qquad d_2=\text{diag}\left(-1,1,-1\right),\\
&&d_3=\text{diag}\left(-1,-1,1\right),\qquad d_4=\text{diag}\left(1,1,1\right)\,.
\end{eqnarray}
It is easy to see that $G_{4}$ is exactly a trivial identity matrix, and we can further check that
\begin{equation}
G^2_{i}=1,\qquad G_{i}G_{j}=G_{j}G_{i}=G_{k}~~\text{with}~~i\neq j\neq k\neq4.
\end{equation}
Hence the residual flavor symmetry of the neutrino mass matrix is $Z_2\times Z_2$ Klein group. On the other hand, given the remnant Klein symmetry in the neutrino sector and the associated 3-dimensional unitary representation matrices, one can straightforwardly construct the diagonalization matrix $U_{\nu}$. Similarly, the remnant flavor symmetry $G_{l}$ of the charged lepton mass matrix satisfies
\begin{equation}
G^{\dagger}_{l}\mathcal{M}_{l}G_{l}=\mathcal{M}_{l}\,.
\end{equation}
Since $\mathcal{M}_{l}$ is diagonal in the chosen basis and the three charged lepton masses are unequal, $G_{l}$ can only be a unitary diagonal matrix, i.e.,
\begin{equation}
G_{l}=\text{diag}\left(e^{i\alpha_{e}}, e^{i\alpha_{\mu}}, e^{i\alpha_{\tau}}\right)\,,
\end{equation}
where $\alpha_{e,\mu,\tau}$ are arbitrary real parameters. Hence the charged lepton mass term generically admits a $U(1)\times U(1)\times U(1)$ remnant flavor symmetry. Conversely, if $G_{l}$ is diagonal with non-degenerate eigenvalues, $\mathcal{M}_{l}$ would be forced to be real. The idea of residual symmetries $G_{\nu}$ and $G_{l}$ arising from some underlying discrete flavor symmetry group $\mathcal{G}_f$ has been extensively explored, and many flavor models have been constructed~\cite{Altarelli:2010gt,Ishimori:2010zr,Grimus:2012dk,King:2013eh,King:2014nza}.

In the following, we shall investigate the remnant CP symmetry of the lepton mass terms. They don't receive enough attention they deserve in the past.  The CP transformation of the left-handed neutrino fields is defined via
\begin{equation}
\nu_{L}(x)\stackrel{CP}{\longmapsto}iX_{\nu}\gamma^{0}C\bar{\nu}^{T}_{L}(x_P)\,,
\end{equation}
where $x_{P}=(t,-\vec{x})$, and $X_{\nu}$ is a $3\times3$ unitary matrices acting on generation space. $X_{\nu}$ is usually called generalized CP transformation in the literature~\cite{Ecker:1981wv,Grimus:1995zi,Branco:2011zb}, since it is an identity matrix in conventional CP transformation. The Lagrangian of the neutrino mass term in Eq.~\eqref{eq:mass_Lagrange} would be invariant if the neutrino mass matrix $m_{\nu}$ fulfills
\begin{equation}
\label{eq:CP_invar_cond}X^{T}_{\nu}m_{\nu}X_{\nu}=m^{\ast}_{\nu}\,.
\end{equation}
With the general form of $m_{\nu}$ in Eq.~\eqref{eq:mnu}, we obtain
\begin{equation}
\left(U^{\dagger}_{PMNS}X_{\nu}U^{\ast}_{PMNS}\right)^{T}m_{diag}\left(U^{\dagger}_{PMNS}X_{\nu}U^{\ast}_{PMNS}\right)=m_{diag}\,,
\end{equation}
which yields
\begin{equation}
\label{eq:remnant_Xi}U^{\dagger}_{PMNS}X_{\nu}U^{\ast}_{PMNS}=\text{diag}\left(\pm1, \pm1, \pm1\right)
\end{equation}
Therefore there are eight possibilities for $X_{\nu}$. However, only four of them are relevant, and they can chosen to be
\begin{equation}
X_{i}=U_{PMNS}\,d_i\,U^{T}_{PMNS},~~ i=1,2,3,4\;.
\end{equation}
The remaining four can be obtained from the above chosen ones by multiplying an over $-1$ factor. Note that we can not distinguish $X_{\nu}$ from $-X_{\nu}$ since the minus sign can be absorbed by redefining the neutrino fields. Moreover, we see that the remnant CP transformations $X_i$ are symmetric unitary matrices:
\begin{equation}
X_i=X^{T}_i\,,
\end{equation}
otherwise the light neutrino masses would be degenerate. The same constraint that the remnant CP transformations in the neutrino sector should be symmetric is also obtained in Ref.~\cite{Feruglio:2012cw} in another way. It is remarkable that the remnant flavor symmetry can be induced by the remnant CP symmetry. From Eq.~\eqref{eq:CP_invar_cond}, it is easy to obtain,
\begin{equation}
X^{\dagger}_jX_{i}^Tm_{\nu}X_iX_j^{\ast}=m_{\nu}\,.
\end{equation}
This means that successively performing two CP transformations $X_iX_j^{\ast}$ is equivalent to a flavor symmetry transformations. Concretely we have the following relations:
\begin{eqnarray}
\nonumber&&X_2X^{\ast}_3=X_3X^{\ast}_2= X_4X^{\ast}_1=X_1X^{\ast}_4=G_1,\\
\nonumber&&X_1X^{\ast}_3=X_3X^{\ast}_1=X_4X^{\ast}_2=X_2X^{\ast}_4=G_2,\\
\nonumber&&X_1X^{\ast}_2=X_2X^{\ast}_1=X_4X^{\ast}_3=X_3X^{\ast}_4=G_3,\\
\label{eq:CP_flavor}&&X_1X^{\ast}_1=X_2X^{\ast}_2=X_3X^{\ast}_3=X_4X^{\ast}_4=G_4=1\,.
\end{eqnarray}
As a consequence, once we impose a set of generalized CP transformations onto the theory, there is always an accompanied flavor symmetry generated. Furthermore, Eq.~\eqref{eq:CP_flavor} implies that any residual CP transformation can be expressed in terms of the remaining ones as follows,
\begin{equation}
\label{eq:CP_3_indep}X_{i}=X_{j}X^{\ast}_{m}X_{n},\qquad i\neq j\neq m\neq n\,.
\end{equation}
In other words, only three of the four remnant CP transformations are independent. In the same fashion, the CP transformation of the charged lepton fields is
\begin{equation}
l_{L}(x)\stackrel{CP}{\longmapsto}iX_{l}\gamma^{0}C\bar{l}^{\,T}_{L}(x_P)\,,
\end{equation}
for the symmetry to hold, the mass matrices $\mathcal{M}_l$ has to satisfy
\begin{equation}
X^{\dagger}_{l}\mathcal{M}_{l}X_{l}=\mathcal{M}^{*}_l
\end{equation}
In the chosen basis where $\mathcal{M}_l$ is diagonal, $X_{l}$ can only be a diagonal phase matrix, i.e.,
\begin{equation}
X_{l}=\text{diag}\left(e^{i\beta_e}, e^{i\beta_{\mu}}, e^{i\beta_{\tau}}\right)\,,
\end{equation}
where $\beta_{e, \mu, \tau}$ are real. In short, the remnant CP symmetry can be constructed from the mixing matrix, and its explicit form can be determined more precisely with the improving measurement accuracy of the mixing angles and CP phases. Before closing this section, we present the above discussed residual CP symmetry in an arbitrary basis:
\begin{equation}
X_{i}=U_{\nu}\,d_i\,U^{T}_{\nu}~~(i=1,2,3,4),\qquad X_{l}=U_{l}\,\text{diag}\left(e^{i\beta_e}, e^{i\beta_{\mu}}, e^{i\beta_{\tau}}\right)U^{T}_{l}\,,
\end{equation}
which can be derived in exactly the same way. In the end, we conclude that the remnant symmetries can be constructed from the mixing matrix which can be measured experimentally. In the following section, we shall demonstrate that the lepton mixing matrix can be constructed from the postulated remnant CP transformations.

\section{\label{sec:reconstruction_PMNS}Reconstruction of lepton mixing matrix from remnant CP symmetries}
\setcounter{equation}{0}

As has been shown in section~\ref{sec:remnant_symmetry}, residual CP symmetries can be derived from mixing matrix, and vice versa lepton mixing matrix can be constructed from the remnant CP symmetries in the neutrino and the charged lepton sectors. In concrete models, we can start from a set of CP transformations $\mathcal{X_{CP}}$ which the Lagrange respects at high energy scale. Subsequently $\mathcal{X_{CP}}$ is spontaneously broken by some scalar fields into different remnant symmetries in the neutrino and the charged lepton sectors. The misalignment between the two remnant symmetries is responsible for the mismatch of the rotations which diagonalize the neutrino and charged lepton matrices, and accordingly the PMNS matrix is generated. The remnant CP symmetries would be assumed hereinafter and we shall not consider how the required vacuum compatible with the remnant symmetries is dynamically achieved, since the resulting lepton mixing pattern is independent of vacuum alignment mechanism and there are generally more than one methods realizing the desired symmetry breaking in practical model building.

As before we still stick to the charged lepton diagonal basis in the following. If four CP transformations $X_{Ri}~(i=1,2,3,4)$ out of $\mathcal{X}_{CP}$ are conserved by the neutrino mass matrix, where the subscript ``$R$'' denotes remnant. In order to be well-defined, $X_{Ri}$ should be unitary matrices and satisfy:
\begin{equation}
X_{Ri}=X^{T}_{Ri},\qquad  X_{Ri}X^{\ast}_{Rj}=X_{Rj}X^{\ast}_{Ri}=X_{Rm}X^{\ast}_{Rn}=X_{Rn}X^{\ast}_{Rm},\qquad \left( X_{Ri}X^{\ast}_{Rj}\right)^2=1\,,
\end{equation}
for $i\neq j\neq m\neq n$. As shown in Eq.~\eqref{eq:CP_flavor}, a $Z_{2} \times Z_{2}$ remnant flavor symmetry is generated with element of the form $X_{Ri}X^{\ast}_{Rj}~(i\neq j)$. It is well-known that the lepton mixing matrix (except the Majorana phases) is fixed by the residual Klein group up to independent permutations of rows and columns. If the residual flavor symmetry originates from a finite flavor symmetry group at high energy, a complete classification of all possible PMNS matrix has been worked out~\cite{Fonseca:2014koa}. An added bonus here is that the Majorana CP phases can also be determined from the postulated remnant CP transformations although they are not constrained by the remnant flavor symmetry at all. If three CP transformations are preserved by the neutrino mass terms, the fourth one can be generated in the way shown in Eq.~\eqref{eq:CP_3_indep}. Hence there are still four residual CP transformations. Given the explicit forms of remnant CP, the lepton mixing matrix can be straightforwardly calculated.

In the following, we shall investigate the most interesting case in which the neutrino mass matrix is invariant under the action of two residual CP transformations $X_{R1}$ and $X_{R2}$. In concrete models, this situation can be realized in two different ways: only $X_{R1}$ and $X_{R2}$ belong to the beginning CP transformations $\mathcal{X_{CP}}$ or $\mathcal{X_{CP}}$ contains all the four remnant CP transformations, but two of them are broken at low energy. To avoid degenerate light neutrino masses, both $X_{R1}$ and $X_{R2}$ should be symmetric unitary matrices. Furthermore, a residual $Z_2$ flavor symmetry is induced with the generator $G_{R}\equiv X_{R1}X^{\ast}_{R2}=X_{R2}X^{\ast}_{R1}$. It is easy to check that the following consistency equations are fulfilled,
\begin{equation}
\label{eq:consistence}X_{R1}G^{\ast}_{R}X^{-1}_{R1}=G_{R},\qquad X_{R2}G^{\ast}_{R}X^{-1}_{R2}=G_{R}\,.
\end{equation}
As we have $X_{R2}=G_{R}X_{R1}$, the neutrino mass matrix would be invariant under $X_{R2}$ if it is invariant under both CP transformation $X_{R1}$ and flavor transformation $G_{R}$. As a consequence, from now on we shall focus on the residual symmetry $X_{R1}$ and $G_{R}$ for convenience. Firstly we note that only one column of the diagonalization matrix $U_{\nu}$, which coincides with $U_{PMNS}$ in the working basis, is fixed by the single $Z_2$ residual flavor symmetry $G_{R}$. This column is exactly the unique eigenvector of $G_{R}$ with eigenvalue $\pm 1$ if $\det(G_R)=\pm 1$~\cite{Ge:2011ih}. Using the freedom of redefining the charged lepton fields (or changing the basis further but keeping $\mathcal{M}_{l}$ still diagonal), each element of this column can always set to be real and non-negative. Hence the column dictated by $G_{R}$ can be parameterized as
\begin{equation}
\label{eq:v1}v_1=\left(
\begin{array}{c}
\cos\varphi            \\
\sin\varphi\cos\phi    \\
\sin\varphi\sin\phi    \\
\end{array}
\right)\,,
\end{equation}
where both $\varphi$ and $\phi$ are real parameters in the interval of $\left[0, \pi/2\right]$. Consequently $G_{R}$ is given by
\begin{equation}
\label{eq:GR}G_{R}=2v_1v^{\dagger}_1-\mathbb{I}_{3\times3}=2v_1v^T_1-\mathbb{I}_{3\times3}\,,
\end{equation}
where $\mathbb{I}_{3\times3}$ denote a three dimensional unit matrix. Given (postulated) the column vector $v_{1}$, the remaining two columns of the PMNS matrix can be obtained from any orthonormal pair of basis vectors $v^{\prime}$ and $v^{\prime\prime}$ in the plane orthogonal to $v_{1}$ by a unitary rotation. Here we choose
\begin{equation}
v^{\prime}=\left(
\begin{array}{c}
\sin\varphi                    \\
-\cos\varphi\cos\phi         \\
-\cos\varphi\sin\phi         \\
\end{array}
\right),\qquad
v^{\prime\prime}=\left(
\begin{array}{c}
   0           \\
 \sin\phi    \\
 -\cos\phi    \\
\end{array}
\right)\,.
\end{equation}
Moreover, since the remnant symmetry can not predict the ordering of the light neutrino mass eigenvalues, the three column vectors of $U_{PMNS}$ can be permuted in any way you want before comparison with experimental data. Therefore the PMNS matrix is of the form:
\begin{eqnarray}
\nonumber U_{PMNS}&=&
\left(
v_1,v^{\prime}, v^{\prime\prime}\right)U_{23}P\\
\label{eq:UPMNS_Gf}&=&
\left(
\begin{array}{ccc}
\cos\varphi       ~&~   \sin\varphi         &      0           \\
\sin\varphi\cos\phi    ~&~   -\cos\varphi\cos\phi    ~&~  \sin\phi    \\
\sin\varphi\sin\phi    ~&~  -\cos\varphi\sin\phi     ~&~    -\cos \phi  \\
\end{array}
\right)U_{23}P\,,
\end{eqnarray}
where $P$ represents any $3\times3$ permutation matrix, and $U_{23}$ is the unitary rotation matrix
\begin{equation}
U_{23}=
\left(
\begin{array}{ccc}
e^{i\beta_1}    ~&~      0     &    0     \\
0    ~&~  \cos\theta e^{i\beta_2}    &  \sin\theta e^{i(\beta_3+\beta_4)}   \\
0     ~&~  -\sin\theta e^{i(\beta_2-\beta_4)}  &  \cos\theta e^{i\beta_3}
\end{array}
\right)\,,
\end{equation}
where the rotation angle $\theta$ and phases $\beta_{1,2,3,4}$ are free real parameters. The values of $\beta_{1,2,3,4}$ would be further constrained by the remnant CP transformation $X_{R1}$.

From Eq.~\eqref{eq:GR}, we see that the residual flavor transformation $G_R$ could be chosen to be real $G_{R}=G^{\ast}_{R}$ without loss of generality. Then the consistency equation of Eq.~\eqref{eq:consistence} implies that $G_{R}X_{R1}=X_{R1}G_{R}$. Hence $v_1$ is a simultaneous eigenvector of $G_{R}$ and $X_{R1}$. Moreover, given an eigenvector $v$ of a symmetric unitary matrix, it is easy to shown that the complex conjugate $v^{\ast}$ is also an eigenvector with the same eigenvalue. As a result, one can always take the eigenvectors of a symmetric unitary matrix to be real. Therefore the remaining two eigenvectors of $X_{R1}$ denoted by $v_{2}$ and $v_{3}$ could be real. They are mutually orthogonal and both orthogonal to $v_1$. Accordingly $v_{2,3}$ can be written as follows,
\begin{eqnarray}
\nonumber v_2&=&\left(
\begin{array}{c}
\sin\varphi\cos\rho   \\
-\sin\phi\sin\rho-\cos\varphi\cos\phi\cos\rho   \\
\cos\phi\sin\rho-\cos\varphi\sin\phi\cos\rho
\end{array}\right)
=v^{\prime}\cos\rho-v^{\prime\prime}\sin\rho\,,\\
\label{eq:v23}v_3&=&\left(
\begin{array}{c}
\sin\varphi\sin\rho     \\
\sin\phi\cos\rho-\cos\varphi\cos\phi\sin\rho    \\
-\cos\phi\cos\rho-\cos\varphi\sin\phi\sin\rho
\end{array}\right)
=v^{\prime}\sin\rho+v^{\prime\prime}\cos\rho  \,,
\end{eqnarray}
where $\rho$ is a real parameter. The remnant CP transformation $X_{R1}$ can be constructed from its eigenvectors $v_{1,2,3}$ via
\begin{equation}
\label{eq:XR1}X_{R1}=e^{i\kappa_1}v_1v^T_1+e^{i\kappa_2}v_2v^T_2+ e^{i\kappa_3}v_3v^T_3\,,
\end{equation}
where $e^{i\kappa_1}$, $e^{i\kappa_2}$ and $e^{i\kappa_3}$ are the eigenvalues of $X_{R1}$. Note that the eigenvalues of a unitary matrix must be complex numbers with modulus one. Then another CP transformation $X_{R2}$ is
\begin{equation}
\label{eq:XR2}X_{R2}=G_{R}X_{R1}=e^{i\kappa_1}v_1v^T_1-e^{i\kappa_2}v_2v^T_2-e^{i\kappa_3}v_3v^T_3\,.
\end{equation}
We see that $v_{1}$, $v_{2}$ and $v_{3}$ are also the eigenvectors of $X_{R2}$ while the corresponding eigenvalues become $e^{i\kappa_1}$, $-e^{i\kappa_2}$ and $-e^{i\kappa_3}$ respectively. Now we turn to explore the constraint of the remnant CP transformation $X_{R1}$. The neutrino mass matrix $m_{\nu}$ invariant under the action of $X_{R1}$ should fulfill $X^{T}_{R1}m_{\nu}X_{R1}=m^{\ast}_{\nu}$ which leads to
\begin{equation}
X_{R1}U^{\ast}_{PMNS}=U_{PMNS}\widehat{X}_{R1}\,,
\end{equation}
as shown in Eq.~\eqref{eq:remnant_Xi}, where $\widehat{X}_{R1}=\text{diag}(\pm1, \pm1, \pm1)$. Inserting the expressions of $X_{R1}$ in Eq.~\eqref{eq:XR1} and $U_{PMNS}$ in Eq.~\eqref{eq:UPMNS_Gf}, we obtain
\begin{equation}
\label{eq:constraint_CP}\text{diag} \left(e^{i\kappa_1}, e^{i\kappa_2}, e^{i\kappa_3}\right)U^{\prime\ast}_{23}P= U^{\prime}_{23}P\widehat{X}_{R1}\,,
\end{equation}
where we have defined
\begin{equation}
U^{\prime}_{23}=\left(
\begin{array}{ccc}
1   ~&~     0    ~&~   0            \\
0   ~&~  \cos\rho ~&~ -\sin \rho    \\
0   ~&~  \sin\rho ~&~ \cos\rho
\end{array}
\right)U_{23}
\end{equation}
Further premultiplying $\text{diag}\left(e^{-i\kappa_1/2}, e^{-i\kappa_2/2},  e^{-i\kappa_3/2} \right)$ and post-multiplying $\widehat{X}^{-1/2}_{R1}P^T$ on both sides of Eq.~\eqref{eq:constraint_CP}, we have
\begin{equation}
\text{diag}(e^{i\frac{\kappa_1}{2}}, e^{i\frac{\kappa_2}{2}}, e^{i\frac{\kappa_3}{2}}) U^{\prime\ast}_{23}(P\widehat{X}^{-\frac{1}{2}}_{R1}P^T)=
\text{diag}(e^{-i\frac{\kappa_1}{2}}, e^{-i\frac{\kappa_2}{2}}, e^{-i\frac{\kappa_3}{2}}) U^{\prime}_{23} (P\widehat{X}^{\frac{1}{2}}_{R1}P^T)\,.
\end{equation}
We can see that the left side of this equation is equal to the complex conjugate of the right side. Notice that $U^{\prime}_{23}$ is a block diagonal matrix and $P\widehat{X}^{1/2}_{R1}P^T$ is diagonal matrix. This equation is satisfied if and only if the combination $\text{diag}(e^{-i\frac{\kappa_1}{2}}, e^{-i\frac{\kappa_2}{2}}, e^{-i\frac{\kappa_3}{2}}) U^{\prime}_{23} (P\widehat{X}^{\frac{1}{2}}_{R1}P^T)$ is a block diagonal real orthogonal matrix, i.e.,
\begin{equation}
\text{diag}\big(e^{-i\frac{\kappa_1}{2}}, e^{-i\frac{\kappa_2}{2}}, e^{-i\frac{\kappa_3}{2}}\big) U^{\prime}_{23}(P\widehat{X}^{\frac{1}{2}}_{R1}P^T)=
\left(
\begin{array}{ccc}
1        ~&~        0       ~&~        0         \\
0        ~&~    \cos\theta  ~&~    \sin\theta   \\
0        ~&~    -\sin\theta  ~&~    \cos\theta    \\
\end{array}
\right)\equiv O_{23}\,,
\end{equation}
where each entry of $O_{23}$ is determined up to a possible minus sign which can always be absorbed into the free parameter $\theta$ and $\widehat{X}_{R1}$. Then we have
\begin{equation}
U^{\prime}_{23}=\text{diag}\big(e^{i\frac{\kappa_1}{2}}, e^{i\frac{\kappa_2}{2}}, e^{i\frac{\kappa_3}{2}}\big)O_{23}(P\widehat{X}^{-\frac{1}{2}}_{R1}P^T)\,.
\end{equation}
Therefore if the neutrino mass term has two remnant CP transformations $X_{R1}$ and $X_{R2}$ given by Eqs.(\ref{eq:XR1},\ref{eq:XR2}) in the charged lepton diagonal basis, then lepton mixing matrix would be constrained to be of the form
\begin{eqnarray}
U_{PMNS}&=&
\left(
\begin{array}{ccc}
\cos\varphi    ~&~ \sin\varphi  ~&~   0    \\
\sin\varphi\cos\phi   ~&~  -\cos\varphi\cos\phi    ~&~  \sin\phi  \\
\sin\varphi\sin\phi   ~&~  -\cos\varphi\sin\phi    ~&~  -\cos\phi  \end{array}
\right)
\left(
\begin{array}{ccc}
1   ~&~   0   ~&~    0           \\
0   ~&~  \cos\rho  ~&~  \sin\rho     \\
0   ~&~  -\sin\rho ~&~  \cos\rho
\end{array}
\right)
\left(
\begin{array}{ccc}
e^{i\frac{\kappa_1}{2}}   &   0   &    0       \\
0 &  e^{i\frac{\kappa_2}{2}}  &  0       \\
0 &  0    &   e^{i\frac{\kappa_3}{2}}
\end{array}
\right)\nonumber\\
\label{eq:PMNS_2CP}&&\times
\left(
\begin{array}{ccc}
1   ~&~  0   ~&~  0     \\
0   ~&~ \cos\theta   ~&~  \sin\theta   \\
0   ~&~ -\sin\theta  ~&~  \cos\theta
\end{array}
\right)P\widehat{X}^{-\frac{1}{2}}_{R1}\,.
\end{eqnarray}
Note that the remnant CP transformations $X_{R1,2}$ depend on six real parameters $\varphi$, $\phi$, $\rho$, $\kappa_1$, $\kappa_2$ and $\kappa_3$, as show in Eqs.(\ref{eq:XR1},\ref{eq:XR2}). The parameters $\kappa_{1,2,3}$ are the arguments of the eigenvalues of $X_{R1,2}$ up to $\pi$ while the eigenvectors are expressed in terms of $\varphi$, $\phi$ and $\rho$. In this approach, $X_{R1,2}$ can take any values you want, and accordingly $\varphi$, $\phi$, $\rho$ and $\kappa_{1,2,3}$ are free input parameters. It is remarkable that the resulting PMNS matrix is independent of how the remnant CP $X_{R1,2}$ is dynamically realized. Once the values of $X_{R1,2}$ (the values of $\varphi$, $\phi$, $\rho$ and $\kappa_{1,2,3}$) are known (postulated), the PMNS matrix compatible with $X_{R1,2}$ would follow immediately from Eq.~\eqref{eq:PMNS_2CP}, and it only depends on one free parameter $\theta$. It is obvious that $\widehat{X}^{-\frac{1}{2}}_{R1}$ is a diagonal matrix with entries $\pm1$ or $\pm i$. Its effect is only shifting the Majorana phases by $\pi$. The permutation matrix $P$ can take six different values. However, exchanging the second and the third columns is equivalent to the redefinitions $\theta\rightarrow\theta-\pi/2$ and $\widehat{X}^{-\frac{1}{2}}_{R1}\rightarrow \text{diag}(1,1,-1)\widehat{X}^{-\frac{1}{2}}_{R1}$. Hence there are only three inequivalent permutations of the columns. In other words, the fixed vector $\left(\cos\varphi, \sin\varphi\cos\phi, \sin\varphi\sin\phi\right)^{T}$ can be the first column, second column of the third column of the PMNS matrix. The phenomenological predictions of the three cases will be discussed in section~\ref{sec:phenomenology}.

Now we consider the last case in which only one remnant CP transformation $X_R$ is preserved by the neutrino mass matrix. Similar to previous cases discussed, $X_R$ should be a symmetric unitary matrix. As a consequence, $X_R$ can be parameterized as \begin{equation}
\label{eq:X_one}X_R=e^{i\kappa_1}v_{1}v^{T}_{1}+e^{i\kappa_2}v_{2}v^{T}_{2}+e^{i\kappa_3}v_{3}v^{T}_{3}\,,
\end{equation}
where $e^{i\kappa_1}$, $e^{i\kappa_2}$ and $e^{i\kappa_3}$ are the eigenvalues of $X_R$, $v_{1}$, $v_{2}$ and $v_{3}$ given by Eqs.(\ref{eq:v1},\ref{eq:v23}) represent its eigenvectors, and they form a set of most general three-dimensional real orthogonal vectors. The invariance of the neutrino mass matrix under $X_R$ leads to the constraint:
\begin{equation}
\label{eq:main}X_RU^{\ast}_{\nu}=U_{\nu}\widehat{X}_R\,,
\end{equation}
where $\widehat{X}_R=\text{diag}\left(\pm1, \pm1, \pm1\right)$ and $U_{\nu}$ denotes the diagonalization matrix of $m_{\nu}$ with $U^{T}_{\nu}m_{\nu}U_{\nu}=\text{diag}(m_1, m_2, m_3)$. This condition can be further simplified into
\begin{equation}
\label{eq:constr_one_CP}\text{diag}\left(e^{i\kappa_1}, e^{i\kappa_2}, e^{i\kappa_3} \right)\widetilde{U}^{\ast}_{\nu}=\widetilde{U}_{\nu}\widehat{X}_R\,,
\end{equation}
where $\widetilde{U}_{\nu}=\left(v_1, v_2, v_3\right)^{\dagger}U_{\nu}$. Premultiplying $\text{diag}\left(e^{-i\kappa_1/2}, e^{-i\kappa_2/2},  e^{-i\kappa_3/2} \right)$ and post-multiplying $\widehat{X}^{-1/2}_{R}$ on both sides of Eq.~\eqref{eq:constr_one_CP}, we obtain
\begin{equation}
\text{diag}(e^{i\frac{\kappa_1}{2}}, e^{i\frac{\kappa_2}{2}}, e^{i\frac{\kappa_3}{2}})\widetilde{U}^{\ast}_{\nu}\widehat{X}^{-\frac{1}{2}}_{R}=\text{diag}(e^{-i\frac{\kappa_1}{2}}, e^{-i\frac{\kappa_2}{2}},e^{-i\frac{\kappa_3}{2}})\widetilde{U}_{\nu}\widehat{X}^{\frac{1}{2}}_{R}\,,
\end{equation}
Therefore the combination $\text{diag}(e^{-i\kappa_1/2}, e^{-i\kappa_2/2},e^{-i\kappa_3/2})\widetilde{U}_{\nu}\widehat{X}^{\frac{1}{2}}_{R}$ is a generic real orthogonal matrix, and it can be expressed as,
\begin{equation}
\text{diag}(e^{-i\frac{\kappa_1}{2}}, e^{-i\frac{\kappa_2}{2}},e^{-i\frac{\kappa_3}{2}})\widetilde{U}_{\nu}\widehat{X}^{\frac{1}{2}}_{R}=O_{3\times3}\,,
\end{equation}
with
\begin{equation}
\label{eq:Orthogonal_matrix}O_{3\times3}=\left(\begin{array}{ccc}
1 & 0 & 0 \\
0 & \cos\theta_1   &   \sin\theta_1 \\
0 & -\sin\theta_1  &   \cos\theta_1
\end{array}\right)
\left(\begin{array}{ccc}
\cos\theta_2   &   0    &    \sin\theta_2 \\
0   &   1   &   0 \\
-\sin\theta_2   &   0   &    \cos\theta_2
\end{array}\right)
\left(\begin{array}{ccc}
\cos\theta_3     &    \sin\theta_3    &    0 \\
-\sin\theta_3    &    \cos\theta_3    & 0   \\
0    &    0     &    1
\end{array}
\right)\,,
\end{equation}
where $\theta_{1,2,3}$ are real parameters, and a possible overall minus sign of $O_{3\times3}$ is dropped since it is insignificant. Hence we conclude that the PMNS matrix, which coincides with $U_{\nu}$ in our working basis, is of the form
\begin{equation}
\label{eq:PMNS_one_CP}U_{PMNS}=\left(v_1, v_2, v_3\right)\text{diag}\big(e^{i\frac{\kappa_1}{2}}, e^{i\frac{\kappa_2}{2}}, e^{i\frac{\kappa_3}{2}}\big)O_{3\times3}\widehat{X}^{-\frac{1}{2}}_{R}\,.
\end{equation}
Comparing with scenario with two residual CP transformations, we still need six input parameters $\varphi$, $\phi$, $\rho$ and $\kappa_{1,2,3}$ to specify the explicit form of the assumed CP transformation, and the resulting lepton mixing matrix $U_{PMNS}$ depends on three free parameters $\theta_{1,2,3}$ instead of one.

\section{\label{sec:phenomenology}Phenomenological implications}
\setcounter{equation}{0}

In this section, we shall present the phenomenological predictions for the lepton mixing parameters when one or two residual CP transformations are conserved by the neutrino mass matrix in the charged lepton diagonal basis. Different independent permutations of the three column vectors of $U_{PMNS}$ would be considered. The possible CP transformations which entail maximal or vanishing Dirac CP violation in the lepton sector are discussed.

\subsection{\label{subsec:1st_column}First column fixed for two remnant CP}

The general expression of $U_{PMNS}$ for two remnant CP is presented in Eq.~\eqref{eq:PMNS_2CP}. The fist column of $U_{PMNS}$ would be completely fixed the imposed remnant symmetry if the permutation matrix $P$ is chosen to be a unit matrix, i.e. $P=\mathbb{I}_{3\times3}$. Then the PMNS matrix is of the form:
\begin{eqnarray}
\nonumber U_{PMNS}&=&
\left(\begin{array}{ccc}
\cos\varphi    ~&~ \sin\varphi   ~&~ 0           \\
\sin\varphi\cos\phi   ~&~ -\cos\varphi\cos\phi   ~&~  \sin\phi    \\
\sin\varphi\sin\phi   ~&~  -\cos\varphi\sin\phi  ~&~-\cos\phi
\end{array}
\right)
\left(\begin{array}{ccc}
1   ~&~   0    ~&~  0       \\
0   ~&~ \cos\rho   ~&~  \sin\rho     \\
0   ~&~  -\sin\rho  ~&~    \cos \rho
\end{array}
\right)\\
\label{eq:PMNS_1st}&&\times
\left(\begin{array}{ccc}
e^{i\frac{\kappa_1}{2}}    &   0   &    0   \\
0  &   e^{i\frac{\kappa_2}{2}} &   0       \\
0  &   0    &    e^{i\frac{\kappa_3}{2}}
\end{array}
\right)
\left(\begin{array}{ccc}
1   ~&~    0     ~&~    0       \\
0   ~&~  \cos\theta  ~&~   \sin\theta  \\
0   ~&~  -\sin\theta  ~&~  \cos\theta
\end{array}\right)\widehat{X}^{-1/2}_{R1}\,.
\end{eqnarray}
For the sake of convenience we introduce $\kappa^{\prime}_2\equiv\kappa_{2}-\kappa_{1}$ and $\kappa^{\prime}_3\equiv\kappa_{3}-\kappa_{1}$. The expressions for the mixing angles and the CP-odd weak basis invariants can be straightforwardly extracted from Eq.~\eqref{eq:PMNS_1st} as follows.
{\small
\begin{eqnarray}
\nonumber\sin^2\theta_{13}&=&\left(\cos^2\rho\sin^2\theta+\sin^2\rho\cos^2\theta+\frac{1}{2}\sin2\rho\sin2\theta
\cos\frac{\kappa^{\prime}_2-\kappa^{\prime}_3}{2}\right)\sin^2\varphi\,,\\
\nonumber\sin^2\theta_{12}&=&\frac{\left(\cos^2\rho\cos^2\theta+\sin^2\rho\sin^2\theta-\frac{1}{2}\sin2\rho\sin2\theta\cos\frac{\kappa^{\prime}_2
-\kappa^{\prime}_3}{2}\right)\sin^2\varphi}{1-\left(\cos^2\rho\sin^2\theta+\sin^2\rho\cos^2\theta+
\frac{1}{2}\sin2\rho\sin2\theta\cos\frac{\kappa^{\prime}_2-\kappa^{\prime}_3}{2}\right)\sin^2\varphi}\,,\\
\nonumber\sin^2\theta_{23}&=&\Big\{\sin^2\phi\sin^2\rho+\cos^2\varphi\cos^2\phi\cos^2\rho-\frac{1}{2}\cos\varphi\sin2\phi\sin2\rho\cos2\theta\\
\nonumber&&-\cos^2\theta\cos2\rho\left(\cos^2\phi\cos^2\varphi-\sin^2\phi\right)\\
\nonumber&&-\frac{1}{2}\cos\frac{\kappa^{\prime}_2-\kappa^{\prime}_3}{2}\left[
\cos\varphi\sin2\phi\cos2\rho+\left(\sin^2\phi-\cos^2\phi\cos^2\varphi\right)\sin2\rho\right]\sin 2\theta\Big\}\Big/\\
\nonumber&&\Big[1-\left(\cos^2\rho\sin^2\theta+\sin^2\rho\cos^2\theta+\frac{1}{2}\sin2\rho\sin2\theta\cos\frac{\kappa^\prime_2-\kappa^\prime_3}{2}\right)
\sin^2\varphi\Big]\,,\\
\nonumber J_{CP}&=&\frac{1}{4}\cos\varphi\sin^2\varphi\sin2\phi\sin2\theta\sin\frac{\kappa^{\prime}_2-\kappa^{\prime}_3}{2}\,,\\
\nonumber |I_1|&=&\frac{1}{4}\sin^22\varphi\left|\cos^2\rho\cos^2\theta\sin\kappa^{\prime}_2+\sin^2\rho\sin^2\theta\sin\kappa^{\prime}_3-
\frac{1}{2}\sin2\rho\sin2\theta\sin\frac{\kappa^{\prime}_2+\kappa^{\prime}_3}{2}\right|\,,\\
\label{eq:mixing_parameters_1st}|I_2|&=&\frac{1}{4}\sin^22\varphi\left|\cos^2\rho\sin^2\theta\sin\kappa^{\prime}_2+\sin^2\rho\cos^2\theta\sin\kappa^{\prime}_3+
\frac{1}{2}\sin2\rho\sin2\theta\sin\frac{\kappa^{\prime}_2+\kappa^{\prime}_3}{2}\right|\,,
\end{eqnarray}}
where $J_{CP}$ is the Jarlskog invariant~\cite{Jarlskog:1985ht},
\begin{eqnarray}
\nonumber J_{CP}&=&\text{Im}\left[\left(U_{PMNS}\right)_{11}\left(U_{PMNS}\right)_{33}\left(U^{\ast}_{PMNS}\right)_{13}\left(U^{\ast}_{PMNS}\right)_{31}\right]\\
&=&\frac{1}{8}\sin2\theta_{12}\sin2\theta_{13}\sin2\theta_{23}\cos\theta_{13}\sin\delta_{CP}\,,
\end{eqnarray}
where $\delta_{CP}$ is the CP-violating Dirac phase in standard parameterization~\cite{pdg}. The invariants $I_1$ and $I_2$ associated with the Majorana phases are defined by~\cite{Branco:1986gr,Jenkins:2007ip,Branco:2011zb},
\begin{eqnarray}
\nonumber I_1&=&\text{Im}\left[\left(U_{PMNS}\right)^2_{12}\left(U_{PMNS}\right)^{\ast2}_{11}\right]=\frac{1}{4}\sin^22\theta_{12}\cos^4\theta_{13}\sin\alpha_{21}\,,\\
I_2&=&\text{Im}\left[\left(U_{PMNS}\right)^2_{13}\left(U_{PMNS}\right)^{\ast2}_{11}\right]=\frac{1}{4}\sin^22\theta_{13}\cos^2\theta_{12}\sin\alpha^{\prime}_{31}\,,
\end{eqnarray}
where $\alpha^{\prime}_{31}\equiv\alpha_{31}-2\delta_{CP}$, $\alpha_{21}$ and $\alpha_{31}$ are the Majorana CP phases~\cite{pdg}. Note that $\alpha^{\prime}_{31}$ enters into the effective mass of the neutrinoless double beta decay. Here both $I_1$ and $I_2$ are presented in terms of absolute values since the sign of $I_1$ and $I_2$ depends on the entry of $\widehat{X}_{R1}$ being $+1$ or $-1$. We see that all the mixing parameters generally depend on both the parameters of CP transformation and the free parameter $\theta$. As the first column of $U_{PMNS}$ is $\left(\cos\varphi, \sin\varphi\cos\phi, \sin\varphi\sin\phi\right)^{T}$ which is dictated by the induced flavor symmetry $G_{R}\equiv X_{R1}X^{\ast}_{R2}=X_{R2}X^{\ast}_{R1}$, solar mixing angle $\theta_{12}$ and reactor mixing angle $\theta_{13}$ are related with each other by
\begin{equation}
\label{eq:theta12_13_1st}\cos^2\theta_{12}\cos^2\theta_{13}=\cos^2\varphi\,.
\end{equation}
Given the global fit results of $3\sigma$ ranges $0.270\leq\sin^2\theta_{12}\leq0.344$ and $0.0188\leq\sin^2\theta_{13}\leq0.0251$~\cite{Gonzalez-Garcia:2014bfa}, the parameter $\varphi$ is constrained to be
\begin{equation}
0.179\pi\leq\varphi\leq0.205\pi\,.
\end{equation}
\begin{figure}[t!]
\begin{center}
\includegraphics[width=1.0\textwidth]{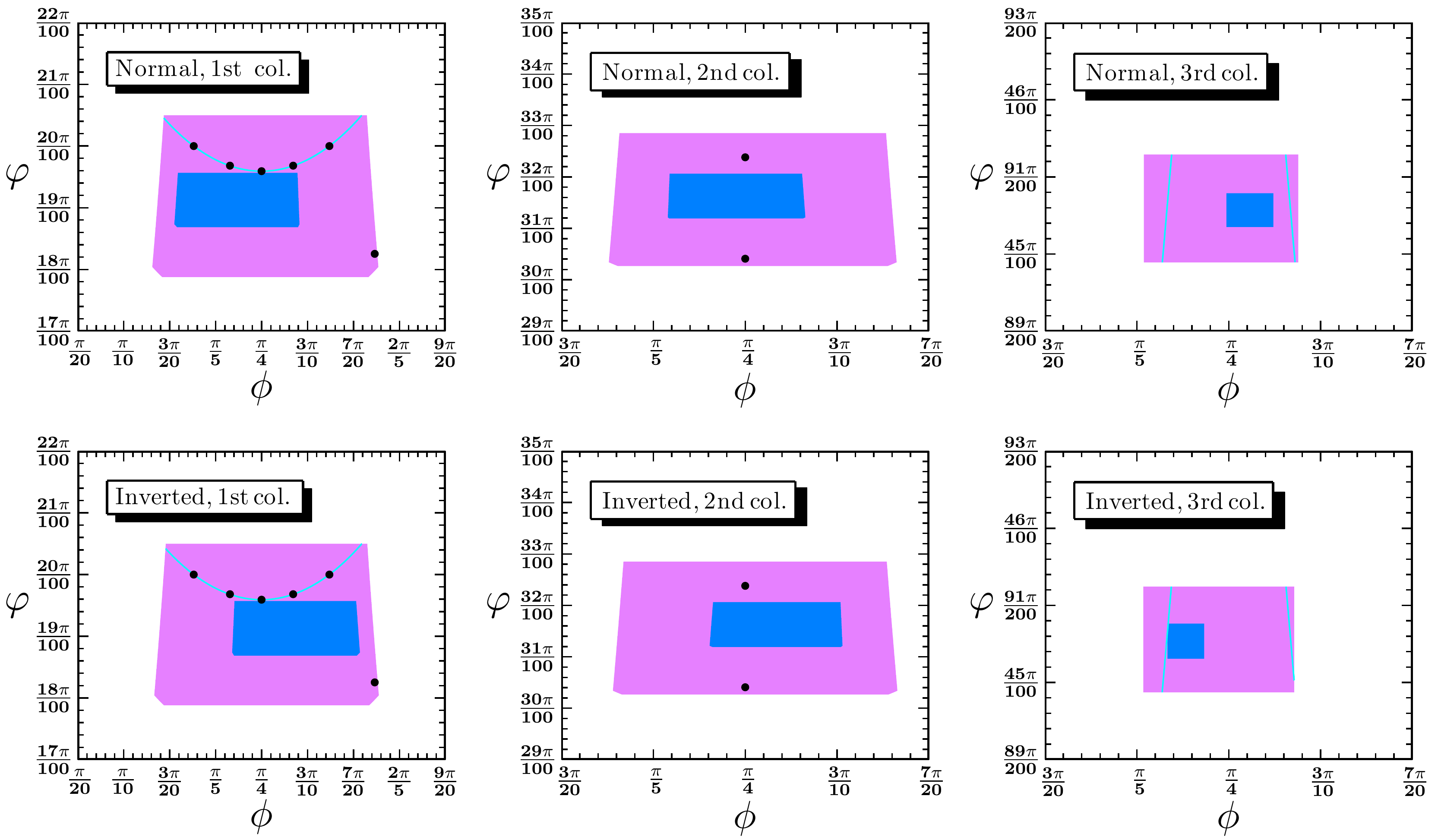}
\caption{\label{fig:range}The allowed regions of $\phi$ and $\varphi$ for mixing parameters in the experimentally preferred $3\sigma$ ranges (pink) and $1\sigma$ ranges (light blue)~\cite{Gonzalez-Garcia:2014bfa}, when two residual CP transformations are conserved by the neutrino mass matrix. The first and second rows are the results for normal ordering and inverted ordering neutrino mass spectrum respectively.
The left column, middle column and the right column are the corresponding results for $\left(\cos\varphi, \sin\varphi\cos\phi, \sin\varphi\sin\phi\right)^{T}$ in the first, second and third column of the $U_{PMNS}$ respectively. The black dots and the green curves denote the viable values of $\phi$ and $\varphi$ given in Eqs.~(\ref{eq:1st_column_finite_flasy}, \ref{eq:2nd_column_finite_flasy}, \ref{eq:3rd_column_finite_flasy}) if the residual flavor symmetry $G_{R}=X_{R1}X^{\ast}_{R2}=X_{R2}X^{\ast}_{R1}$ arises from a finite flavor symmetry group. Note that the cyan curves are the results for the cases that the first or the third column of the PMNS matrix are the corresponding ones of the infinite series $\mathcal{C}_2$ in Eq.~\eqref{eq:C2}.}
\end{center}
\end{figure}
The allowed regions of the parameters $\varphi$ and $\phi$ are displayed in Fig.~\ref{fig:range} when both mixing angles and $\delta_{CP}$ vary in their $3\sigma$ (or $1\sigma$) intervals~\cite{Gonzalez-Garcia:2014bfa}. We see that the $1\sigma$ ranges for normal ordering (NO) and inverted ordering (IO) mass spectrums are different although the corresponding $3\sigma$ results can hardly be distinguished. Moreover, the $1\sigma$ regions are drastically shrunk compared with the $3\sigma$ ones. Therefore more precisely measurement of the mixing parameters can help to eventually pin down the values of $\varphi$ and $\phi$. Furthermore, from $\left|\left(U_{PMNS}\right)_{21}/\left(U_{PMNS}\right)_{31}\right|^2=\cot^2\phi$ we can find a correlation among $\delta_{CP}$ and mixing angles as follows  \begin{equation}
\label{eq:cosd1}\cos\delta_{CP}=\frac{\cos2\theta_{23}(-\sin^2\theta_{12}+\cos^2\theta_{12}\sin^2\theta_{13})+
\cos2\phi(\sin^2\theta_{12}+\cos^2\theta_{12}\sin^2\theta_{13})}{\sin2\theta_{12}\sin2\theta_{23}\sin\theta_{13}}\,.
\end{equation}
Given an input value of $\phi$, we can predict the Dirac CP phase $\delta_{CP}$ from the experimentally measured values of the mixing angles. The  regions of $\cos\delta_{CP}$ with respect to $\phi$ are plotted in Fig.~\ref{fig:phi-cosd}. It is remarkable that $\cos\delta_{CP}$ determined by Eq~\eqref{eq:cosd1} is larger than 1 or smaller than $-1$ for some values of $\phi$, and $\cos\delta_{CP}$ is restricted to be in a rather narrow strip region if all mixing parameters are required to vary in the $1\sigma$ interval. The improved accuracy of the mixing angles will facilitate the determination of $\delta_{CP}$ in this approach.

\begin{figure}[t!]
\begin{center}
\includegraphics[width=1.0\textwidth]{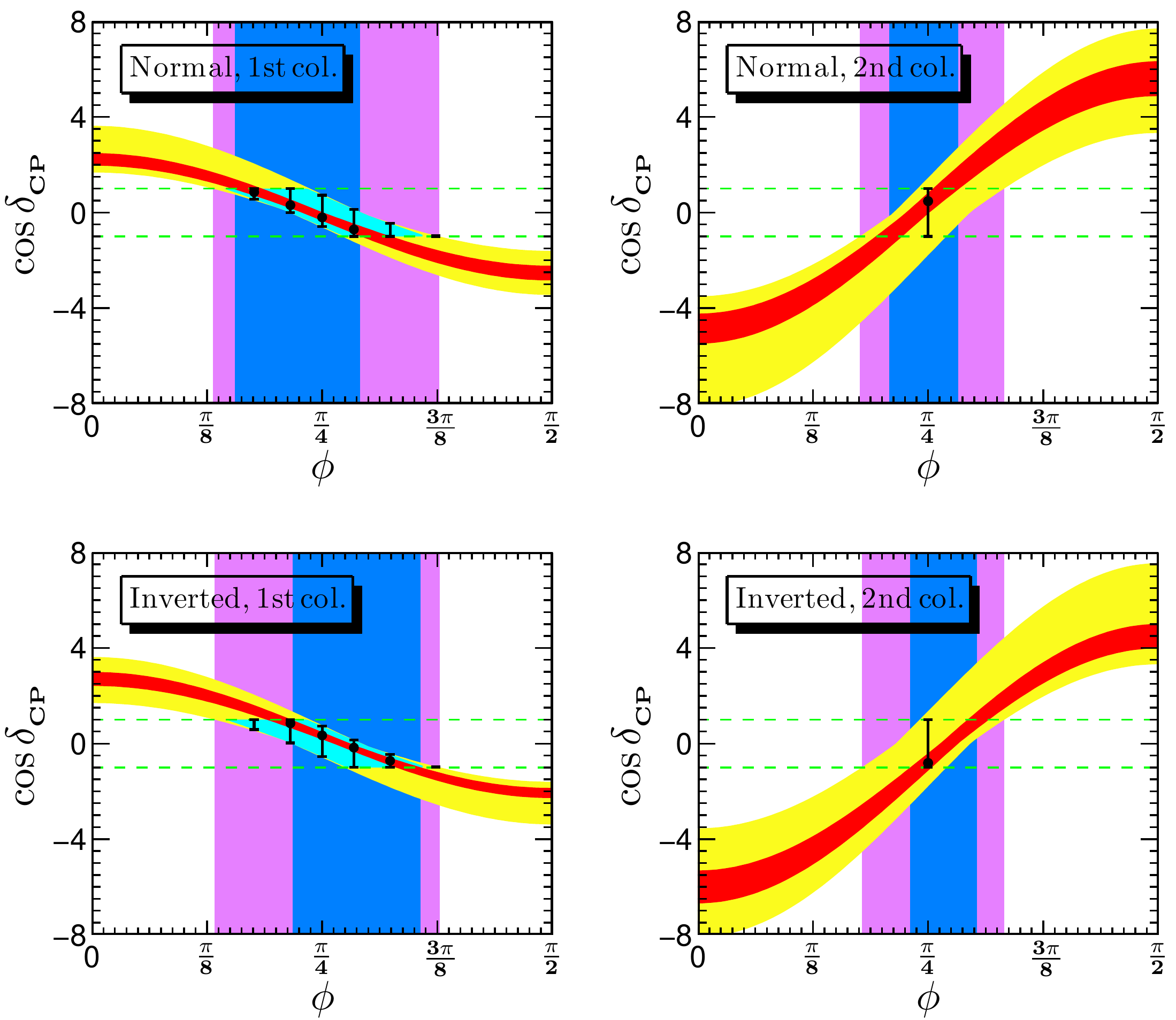}
\caption{\label{fig:phi-cosd}The range of $\cos\delta_{CP}$ versus $\phi$ predicted by the relations of Eqs.(\ref{eq:cosd1},\ref{eq:cosd2}) when all three mixing angles vary in the experimentally preferred $3\sigma$ regions (red) and $1\sigma$ regions (yellow). The constraint of $|\cos\delta_{CP}|\leq1$ isn't included for the time being in order to see clearly the viable $\phi$ values. The panels on the left side and on the right side correspond to the vector $(\cos\varphi, \sin\varphi\cos\phi, \sin\varphi\sin\phi)^{T}$ in the first column and the second column of $U_{PMNS}$ respectively. The panels in the first row and the second row are for normal ordering and inverted ordering mass spectrum respectively. The pink and light blue bands stand for the $3\sigma$ and $1\sigma$ ranges of $\phi$ displayed in Fig.~\ref{fig:range}. The horizontal dashed lines represents the boundaries of $\cos\delta_{CP}=+1$ and $\cos\delta_{CP}=-1$.
The vertical black lines denote the ranges of $\cos\delta_{CP}$ for the discrete $\phi$ and $\varphi$ given in Eqs.~(\ref{eq:1st_column_finite_flasy}, \ref{eq:2nd_column_finite_flasy}). The cyan areas are for the case that the first column of $U_{PMNS}$ is the corresponding one of the infinite series $\mathcal{C}_2$ (the last one of Eq.~\eqref{eq:1st_column_finite_flasy}). Note that the cyan regions nearly overlap with the corresponding $3\sigma$ yellow part.}
\end{center}
\end{figure}

\begin{figure}[t!]
\begin{center}
\begin{tabular}{cc}
\includegraphics[width=0.46\linewidth]{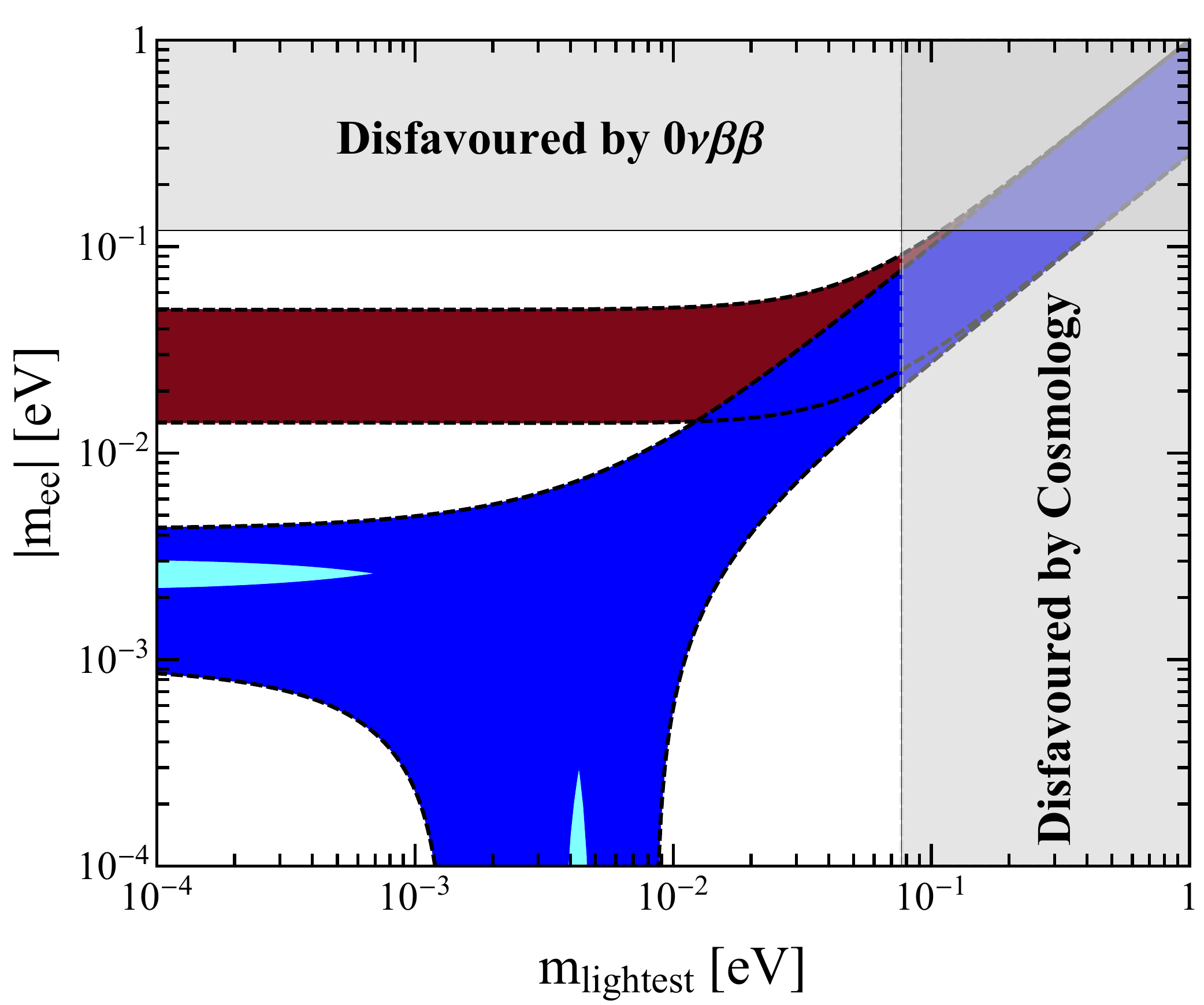} &
\includegraphics[width=0.46\linewidth]{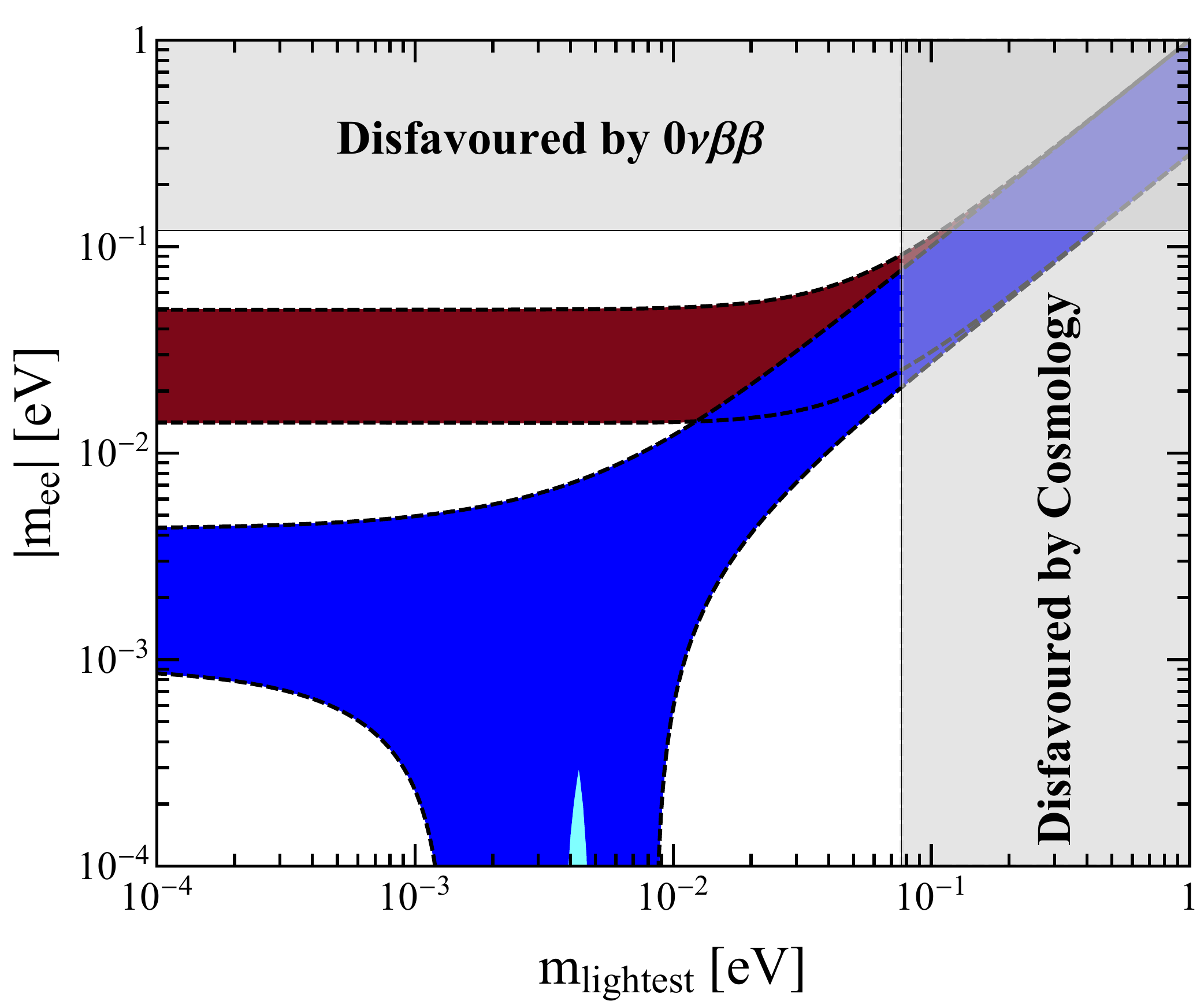} \\
(a)  &  (b)  \\
   &   \\ [-0.16in]
\includegraphics[width=0.46\linewidth]{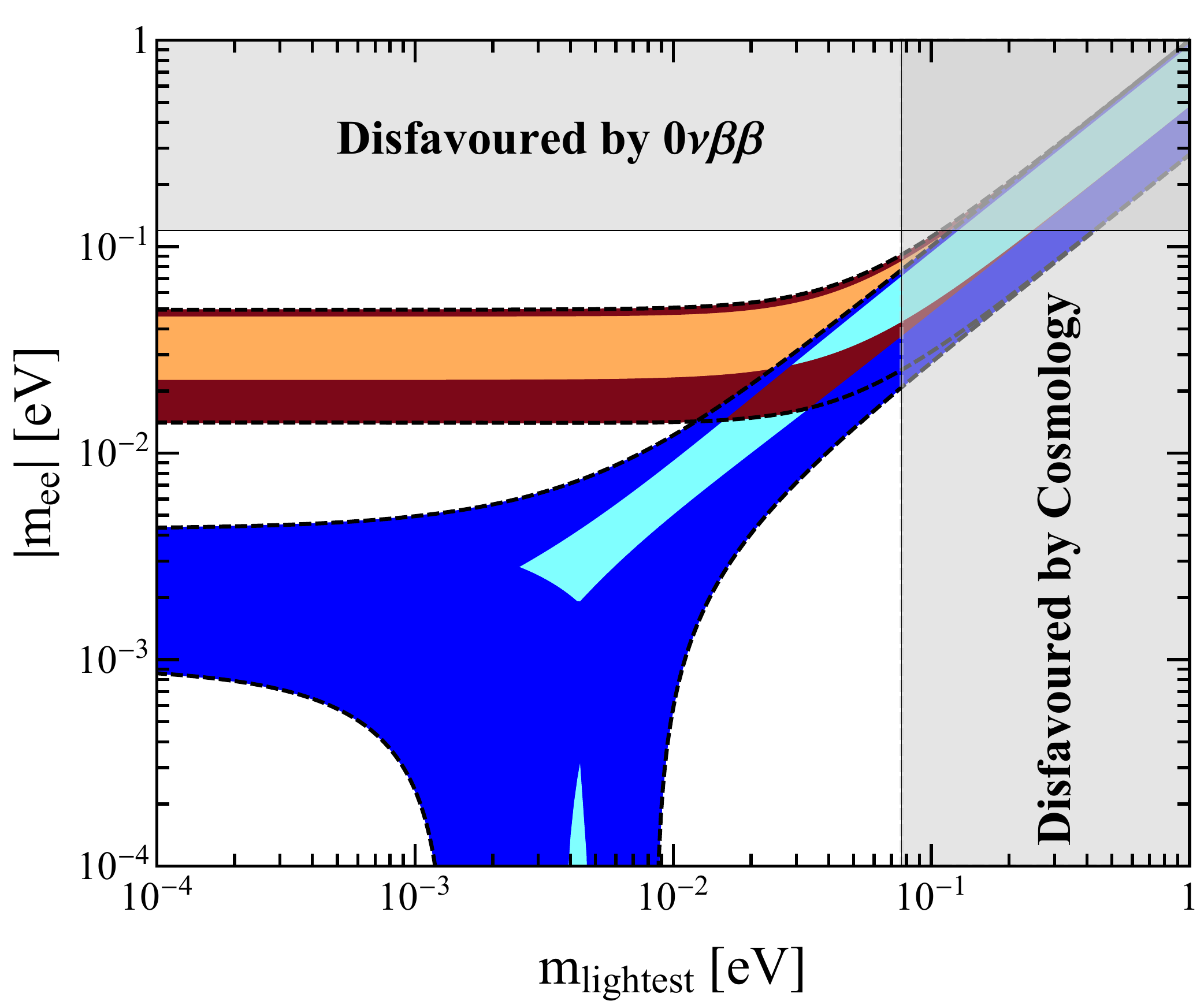} &
\includegraphics[width=0.46\linewidth]{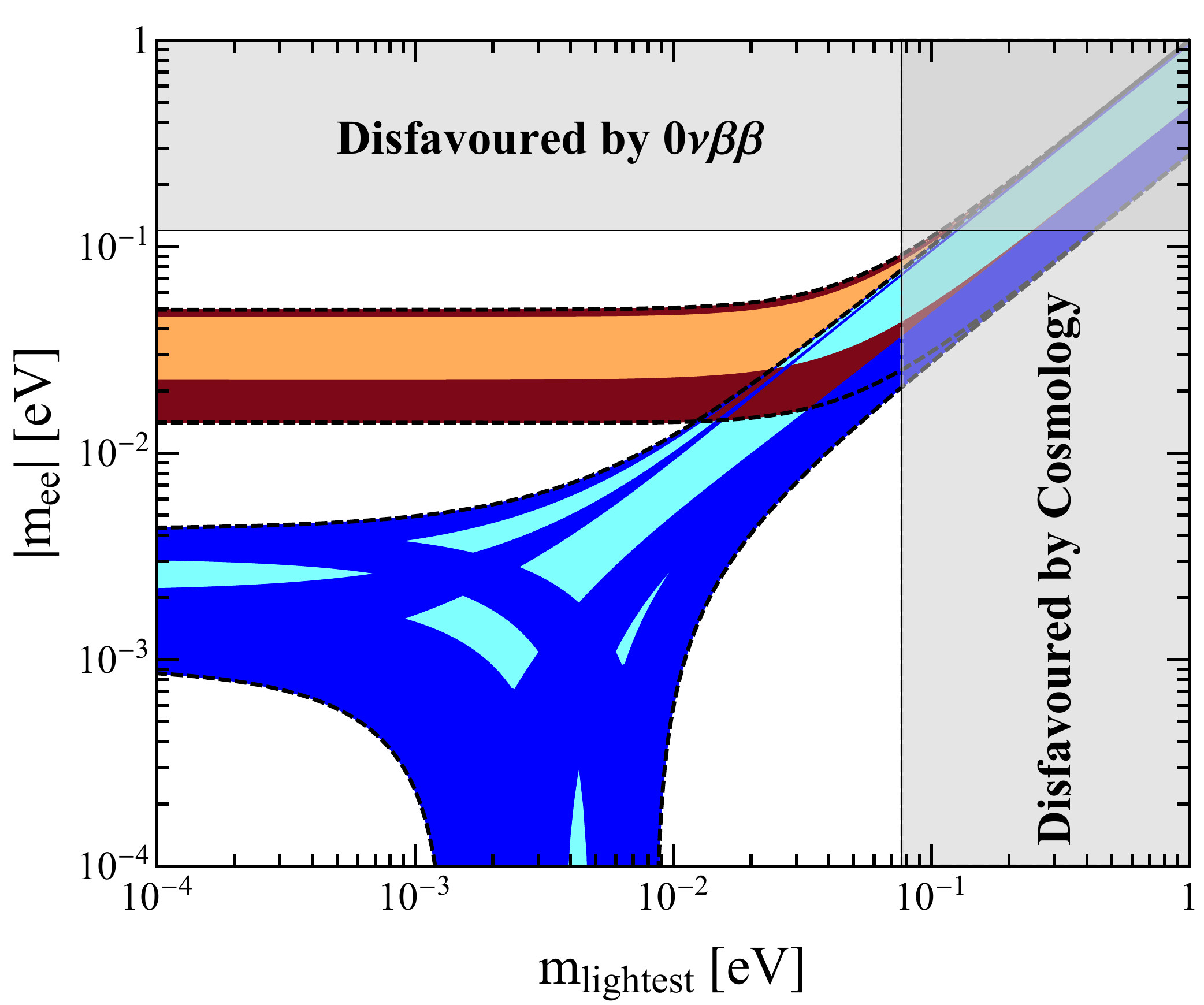} \\
(c)  &  (d)
\end{tabular}
\caption{\label{fig:mee}The $(\beta\beta)_{0\nu}-$decay effective mass $|m_{ee}|$ as a function of the lightest neutrino mass when the conditions of  maximal Dirac phase in Eqs.~(\ref{eq:maximal_1st}, \ref{eq:maximal_3rd}, \ref{eq:maximal_1CP}) are fulfilled. The corresponding mixing parameters are given by Eqs.~(\ref{eq:PMNS_1st_maximal}, \ref{eq:PMNS_2nd_maximal}, \ref{eq:PMNS_3rd_maximal}, \ref{eq:maximal_single}). To be specific, panel (a) is for the case of $\sin(\alpha_{21}-\alpha_{31})=0$, (b) for $\sin\alpha_{31}=0$, (c) for $\sin\alpha_{21}=0$, and (d) for $\sin\alpha_{21}=\sin\alpha_{31}=0$. The cyan and orange areas represent the currently allowed $3\sigma$ regions for normal ordering and inverted ordering mass spectrum respectively~\cite{Gonzalez-Garcia:2014bfa}. The purple and blue regions are theoretical predictions for different scenarios mentioned above. Measurements of EXO-200~\cite{Auger:2012ar,Albert:2014awa} in combination with KamLAND-ZEN~\cite{Gando:2012zm} give a bound of $|m_{ee}|<0.120$ eV. The upper limit on the mass of the lightest neutrino is derived from the latest Planck result $m_1+m_2+m_3<0.230$ eV at $95\%$ confidence level~\cite{Ade:2013zuv}.}
\end{center}
\end{figure}

From the expression of Jarlskog invariant $J_{CP}$ in Eq.~\eqref{eq:mixing_parameters_1st}, we see that $J_{CP}$ vanishes such that the Dirac CP is conserved with $\delta_{CP}=0, \pi$ if $\kappa_{2}=\kappa_{3}$ or $\theta=0, \pi/2, \pi, 3\pi/2$. Notice that both $\varphi$ and $\phi$ can not be equal to $0$ or $\pi/2$ otherwise at least one entry of $U_{PMNS}$ would be zero. Inspired by the indication of maximal $\delta_{CP}$ from T2K Collaboration~\cite{Abe:2013hdq}, we shall explore the condition of maximal Dirac CP-violation. By straightforward but cumbersome calculations, we find that $\cos\delta_{CP}=0$ can be achieved for the parameters
\begin{equation}
\label{eq:maximal_1st}\kappa_3=\kappa_2\pm\pi,\qquad \phi=\frac{\pi}{4},\qquad
\rho=0,\frac{\pi}{2}, \pi~\text{or}~\frac{3\pi}{2}\,,
\end{equation}
no matter what value $\theta$ takes. The corresponding residual CP transformations are of the form
\begin{eqnarray}
\nonumber X_{R1}&=&e^{i\kappa_{+}}\left(
\begin{array}{ccc}
 \cos\frac{\kappa^{\prime}_2}{2}-i\cos2\varphi\sin\frac{\kappa^{\prime}_2}{2}  & -\frac{i}{\sqrt{2}}\sin2\varphi\sin\frac{\kappa^{\prime}_2}{2} & ~-\frac{i}{\sqrt{2}}\sin2\varphi\sin\frac{\kappa^{\prime}_2}{2} \\
 -\frac{i}{\sqrt{2}}\sin2\varphi\sin\frac{\kappa^{\prime}_2}{2} &
   -i\sin^2\varphi\sin\frac{\kappa^{\prime}_2}{2} & \cos\frac{\kappa^{\prime}_2}{2}+i\cos^2\varphi\sin\frac{\kappa^{\prime}_2}{2}\\
 -\frac{i}{\sqrt{2}}\sin2\varphi\sin\frac{\kappa^{\prime}_2}{2} &
   \cos\frac{\kappa^{\prime}_2}{2}+i\cos^2\varphi\sin\frac{\kappa^{\prime}_2}{2}  & -i\sin^2\varphi\sin\frac{\kappa^{\prime}_2}{2}\\
\end{array}
\right),\\
X_{R2}&=&e^{i\kappa_{+}}\left(
\begin{array}{ccc}
\cos2\varphi\cos\frac{\kappa^{\prime}_2}{2}-i\sin\frac{\kappa^{\prime}_2}{2}  & \frac{1}{\sqrt{2}} \sin2\varphi\cos\frac{\kappa^{\prime}_2}{2} &  \frac{1}{\sqrt{2}} \sin2\varphi\cos\frac{\kappa^{\prime}_2}{2}\\
\frac{1}{\sqrt{2}} \sin2\varphi\cos\frac{\kappa^{\prime}_2}{2}  & \sin^2\varphi\cos\frac{\kappa^{\prime}_2}{2}  & -\cos^2\varphi\cos\frac{\kappa^{\prime}_2}{2}-i\sin\frac{\kappa^{\prime}_2}{2} \\
\frac{1}{\sqrt{2}} \sin2\varphi\cos\frac{\kappa^{\prime}_2}{2} & -\cos^2\varphi\cos\frac{\kappa^{\prime}_2}{2}-i\sin\frac{\kappa^{\prime}_2}{2}  & \sin^2\varphi\cos\frac{\kappa^{\prime}_2}{2}\\
\end{array}
\right)\,,
\end{eqnarray}
in case of $\rho=0~\text{or}~\pi$, $\kappa_3=\kappa_2\pm\pi$ and $\phi=\pi/4$, where $\kappa_{+}=(\kappa_1+\kappa_2)/2$. The above formulas for $X_{R1}$ and $X_{R2}$ are interchanged for the remaining values of $\rho=\pi/2~\text{or}~3\pi/2$, $\kappa_3=\kappa_2\pm\pi$ and $\phi=\pi/4$. In this occasion, the lepton mixing parameters are predicted to be
\begin{eqnarray}
\nonumber&\sin^2\theta_{12}=\frac{\cos^2\theta\sin^2\varphi}{1-\sin^2\theta\sin^2\varphi},\qquad \sin^2\theta_{13}=\sin^2\theta\sin^2\varphi,\\
\label{eq:PMNS_1st_maximal}&\sin^2\theta_{23}=\frac{1}{2},\qquad \cos\delta_{CP}=0,\qquad \tan\alpha_{21}=\tan\alpha_{31}=\tan\kappa^\prime_2\,.
\end{eqnarray}
We see that both $\theta_{23}$ and $\delta_{CP}$ are maximal, and the Majorana CP phases $\alpha_{21}$ and $\alpha_{31}$ are equal up to $\pi$. Note that the requirement of Eq.~\eqref{eq:maximal_1st} is a sufficient but not a necessary condition of maximal Dirac CP violation. Since generally $\delta_{CP}$ depends on both the input parameters $\varphi$, $\phi$, $\rho$, $\kappa_{1,2,3}$ associated with the residual CP transformations and the free parameter $\theta$, $\cos\delta_{CP}=0$ can also be achieved for some specific value of $\theta$ (not for any value of $\theta$) even if the condition in Eq.~\eqref{eq:maximal_1st} is not fulfilled. This point can be clearly seen from Fig.~\ref{fig:phi-cosd}. The key observable for Majorana phases is the neutrinoless double beta ($(\beta\beta)_{0\nu}-$) decay. The dependence of the $(\beta\beta)_{0\nu}-$decay amplitude on the neutrino mixing parameters is represented by effective mass $|m_{ee}|$~\cite{pdg}:
\begin{equation}
\label{eq:mee}\left|m_{ee}\right|=\left|m_1\cos^2\theta_{12}\cos^2\theta_{13}+m_2\sin^2\theta_{12}\cos^2\theta_{13}e^{i\alpha_{21}}+m_3\sin^2\theta_{13}e^{i(\alpha_{31}-2\delta_{CP})}\right|\,.
\end{equation}
The effective mass $|m_{ee}|$ for the predicted patterns in Eq.~\eqref{eq:PMNS_1st_maximal} is illustrated in Fig.~\ref{fig:mee}(a). We
see that almost all possible values of $|m_{ee}|$ allowed by experimental data at $3\sigma$ level can be reproduced in this case except a quite small portion in case of NO. Therefore it should be very challenging to testify this texture in $(\beta\beta)_{0\nu}-$decay.

The framework put forward above is very general. $\varphi$ and $\phi$ are free input parameters, and they can take any values. Now we consider another interesting scenario in which the induced flavor symmetry $G_{R}=X_{R1}X^{\ast}_{R2}=X_{R2}X^{\ast}_{R1}$ arises from some finite flavor symmetry group. All possible PMNS matrix has been derived if the residual flavor symmetry is the full $Z_2\times Z_2$ Klein group in the neutrino sector~\cite{Fonseca:2014koa}. The complete list of the lepton mixing matrices contain 17 sporadic $|U_{PMNS}|^2$ patterns and one infinite series denoted by $\mathcal{C}_2$ with
\begin{eqnarray}
\nonumber |U_{PMNS}|^2&=&\frac{1}{3}\left(\begin{array}{ccc}
1+\textrm{Re}\,\sigma & ~1~ & 1-\textrm{Re}\,\sigma\\
1+\textrm{Re}\left(\omega^2 \sigma\right) & ~1~ &
1-\textrm{Re}\left(\omega^2 \sigma\right)\\
1+\textrm{Re}\left(\omega\sigma\right) & ~1~ &
1-\textrm{Re}\left(\omega\sigma\right)
\end{array}\right)\\
\label{eq:C2}&=&\frac{1}{3}\left(
\begin{array}{ccc}
2\cos^2\theta_{\nu}  & ~1~ & 2 \sin^2\theta_{\nu}  \\
2\cos^2\left(\theta_{\nu}-\frac{\pi}{3}\right) & ~1~ &  2\sin^2\left(\theta_{\nu}-\frac{\pi}{3}\right) \\
2\cos^2\left(\theta_{\nu}+\frac{\pi}{3}\right) & ~1~ & 2 \sin^2\left(\theta_{\nu}+\frac{\pi}{3}\right)
\end{array}
\right)\,,
\end{eqnarray}
where $\omega=e^{2i\pi/3}$ and $\sigma=e^{2i\pi p/n}$ is any root of unity, and in the second step we have denoted $\theta_{\nu}\equiv\pi p/n$ for simplicity of notation. In the present context the first column vector $\left(\cos\varphi, \sin\varphi\cos\phi, \sin\varphi\sin\phi\right)^{T}$ of Eq.~\eqref{eq:PMNS_1st} dictated by the residual flavor symmetry $G$ can be any column of the PMNS matrices predicted in Ref.~\cite{Fonseca:2014koa}, once $G$ is assumed to originate from some underlying finite flavor symmetry group. By examining all the possible $|U_{PMNS}|^2$ predicted in Ref.~\cite{Fonseca:2014koa} and considering permutations of rows and columns, we find that only the following forms are compatible with the present data at $3\sigma$ level,
\begin{eqnarray}
\nonumber\left(\begin{array}{c}
\cos^2\varphi\\
\sin^2\varphi\cos^2\phi\\
\sin^2\varphi\sin^2\phi
\end{array}
\right)&=&\frac{1}{8}\left(\begin{array}{c}
3+\sqrt{5}\\
2\\
3-\sqrt{5}
\end{array}
\right),\quad \left(\begin{array}{c}
r_1\\
r_2\\
r_3
\end{array}
\right),\quad \frac{1}{6}\left(\begin{array}{c}
4\\
1\\
1
\end{array}
\right),\quad \left(\begin{array}{c}
r_1\\
r_3\\
r_2
\end{array}
\right),\\
\label{eq:1st_column_finite_flasy}&&\frac{1}{8}\left(\begin{array}{c}
3+\sqrt{5}\\
3-\sqrt{5}\\
2
\end{array}
\right),\quad \frac{1}{8}\left(\begin{array}{c}
3+\sqrt{7}\\
3-\sqrt{7}\\
2
\end{array}
\right),\quad \frac{2}{3}\left(
\begin{array}{c}
\cos^2\theta_{\nu}    \\
\cos^2\left(\theta_{\nu}-\frac{\pi}{3}\right)\\
\cos^2\left(\theta_{\nu}+\frac{\pi}{3}\right)
\end{array}
\right)\,,
\end{eqnarray}
where the parameter $\theta_{\nu}$ should be in the interval of $-0.0647\pi\leq\theta_{\nu}\leq0.0626\pi$ to achieve agreement with the experimental data. $r_{1,2,3}$ are the roots of the equation $56x^3-56x^2+14x-1=0$, and their approximate numerical values are $r_{1}\simeq0.664$, $r_2\simeq0.204$, $r_{3}\simeq0.132$~\cite{Fonseca:2014koa}. Notice that the third column of $\mathcal{C}_2$ is also viable, nevertheless it is related to the first column of $\mathcal{C}_2$ (the last one in Eq.~\eqref{eq:1st_column_finite_flasy}) via $\theta_{\nu}$ redefinition. Moreover, the former five column vectors are of the same form as the last one with
$\theta_{\nu}=\arccos\frac{\sqrt{6}+\sqrt{30}}{8}$, $\frac{1}{6}\arctan\frac{3\sqrt{3}}{13}$, 0, $\pi-\frac{1}{6}\arctan\frac{3\sqrt{3}}{13}$ and $\pi-\arccos\frac{\sqrt{6}+\sqrt{30}}{8}$ respectively.
The values of $\varphi$ and $\phi$ for the quantized columns in Eq.~\eqref{eq:1st_column_finite_flasy} are plotted in Fig.~\ref{fig:range}. Utilizing Eqs.(\ref{eq:theta12_13_1st}, \ref{eq:theta12_13_2nd}) and Eqs.(\ref{eq:cosd1}, \ref{eq:cosd2}), we can predict the values of $\cos\delta_{CP}$ and $\sin^2\theta_{12}$ when the mixing angles vary within their $3\sigma$ ranges. These results are collected in Table~\ref{tab:fixed_column}. Obviously precise measurement of $\theta_{12}$ and the Dirac CP phase $\delta_{CP}$ can test this scenario. Note that $\theta_{12}$ can be measured with quite good accuracy by JUNO experiment~\cite{JUNO}.

\begin{table}[t!]
\renewcommand{\arraystretch}{1.2}
\begin{center}
\begin{tabular}{|c|c|c|c|}\hline\hline
  &  $(\cos^2\varphi, \sin^2\varphi\cos^2\phi, \sin^2\varphi\sin^2\phi)^{T}$  &   $\cos\delta_{CP}$  &   $\sin^2\theta_{12}$  \\\hline
\multirow{7}{*}{1st col.}  &$\frac{1}{8}(3+\sqrt{5}, 2, 3-\sqrt{5})^T\in {\cal CD}_{3}, {\cal CD}_{4}$ & $0.572\rightarrow1$  & $0.329\rightarrow 0.333$ \\\cline{2-4}

  & $(r_1,r_2,r_3)^T\in{\cal C}_{4}$ &  $0.0124\rightarrow1$ & $0.319\rightarrow 0.323$ \\\cline{2-4}

  & $\frac{1}{6}(4, 1, 1)^T\in{\cal C}_{5}, {\cal C}_{9}$ &  $-0.568\rightarrow0.723$  & $0.316\rightarrow 0.321$ \\\cline{2-4}

  & $(r_1,r_3,r_2)^T\in{\cal C}_{4}$ &  $-1\rightarrow0.134$  & $0.319\rightarrow 0.323$ \\\cline{2-4}

 & $\frac{1}{8}(3+\sqrt{5}, 3-\sqrt{5}, 2)^T\in{\cal CD}_{3}, {\cal CD}_{4}$  & $-1\rightarrow-0.454$ & $0.329\rightarrow 0.333$ \\\cline{2-4}

  & $\frac{1}{8}(3+\sqrt{7}, 3-\sqrt{7}, 2)^T\in{\cal C}_{3}$ &  $-1\rightarrow-0.969$  & $0.276\rightarrow0.281$\\\cline{2-4}

 &   &     &     \\ [-0.19in]
 & $\begin{array}{c}
 \frac{2}{3}\left(\cos^2\theta_{\nu}, \cos^2\left(\theta_{\nu}-\frac{\pi}{3}\right), \cos^2\left(\theta_{\nu}+\frac{\pi}{3}\right)\right)^{T}\in\mathcal{C}_2,\\
 -0.0647\pi\leq\theta_{\nu}\leq0.0626\pi
 \end{array} $  & $-1\rightarrow 1$  &  $0.316\rightarrow0.346$  \\\hline

 \multirow{2}{*}{2nd col.} & $\frac{1}{3}(1, 1, 1)^T\in{\cal C}_{3}, {\cal C}_{30}$ &
$-1\rightarrow1$ & $0.340\rightarrow 0.342$ \\\cline{2-4}

 & $\frac{1}{20}(10-2\sqrt{5}, 5+\sqrt{5}, 5+\sqrt{5})^T\in{\cal C}_{8}, {\cal C}_{11}, {\cal C}_{13}\, {\cal C}_{17}$  &  $-1\rightarrow1$  &  $0.282\rightarrow 0.284$ \\\hline\hline

\end{tabular}
\end{center}
\renewcommand{\arraystretch}{1.0}
\caption{\label{tab:fixed_column} The predictions for $\cos\delta_{CP}$ and $\sin^2\theta_{12}$ by Eqs.~(\ref{eq:cosd1}, \ref{eq:cosd2}) and Eqs.(\ref{eq:theta12_13_1st}, \ref{eq:theta12_13_2nd}) when the vector $\left(\cos\varphi, \sin\varphi\cos\phi, \sin\varphi\sin\phi\right)^{T}$ is placed in the first or the second column of the PMNS matrix and mixing angles vary in the experimentally preferred $3\sigma$ ranges~\cite{Gonzalez-Garcia:2014bfa}. Here we assume that the induced residual flavor symmetry $G_{R}=X_{R1}X^{\ast}_{R2}=X_{R2}X^{\ast}_{R1}$ originates from a finite flavor symmetry group. Consequently the fixed column can be any column of the PMNS matrices listed in Ref.~\cite{Fonseca:2014koa}. $\mathcal{C}_i$ and $\mathcal{CD}_i$ are notations for mixing patterns introduced in~\cite{Fonseca:2014koa}.}
\end{table}

\subsection{\label{subsec:2nd_column}Second column fixed for two remnant CP}

The column permutation $P$ takes the value
\begin{equation}
P=\left(\begin{array}{ccc}
0  &  1   &  0    \\
0  &  0   &  1    \\
1  &  0   &  0     \\
\end{array}
\right)
\end{equation}
The PMNS matrix is given by
\begin{eqnarray}
\nonumber U_{PMNS}&=&\left(
\begin{array}{ccc}
  0   ~&~  \cos\varphi   ~&~   \sin\varphi    \\
 \sin\phi  ~&~  \sin\varphi\cos\phi   ~&~  -\cos\varphi\cos\phi   \\
-\cos\phi  ~&~  \sin\varphi\sin\phi   ~&~  -\cos\varphi\sin\phi
\end{array}\right)
\left(\begin{array}{ccc}
\cos \rho   ~&~  0   ~&~  -\sin\rho   \\
0   ~&~  1   ~&~   0           \\
\sin\rho   ~&~   0  ~&~  \cos\rho
\end{array}
\right)\\
&&\times\left(
\begin{array}{ccc}
e^{i\frac{\kappa_3}{2}}   &    0   &    0    \\
0   &   e^{i\frac{\kappa_1}{2}}  &   0     \\
0   &   0    &    e^{i\frac{\kappa_2}{2}}
\end{array}\right)
\left(\begin{array}{ccc}
\cos\theta   ~&~  0   ~&~  -\sin\theta    \\
0   ~&~   1   ~&~    0      \\
\sin\theta   ~&~  0   ~&~  \cos\theta
\end{array}
\right)\widehat{X}^{-1/2}_{R1}\,.
\end{eqnarray}
We see that the second column is $\left(\cos\varphi, \sin\varphi\cos\phi, \sin\varphi\sin\phi\right)^{T}$ which is irrelevant to $\theta$. The lepton mixing parameters are determined to be
\begin{eqnarray}
\nonumber\sin^2\theta_{13}&=&\left(\cos^2\rho\cos^2\theta+\sin^2\rho\sin^2\theta-\frac{1}{2}\sin2\rho\sin2\theta
\cos\frac{\kappa^{\prime}_2-\kappa^{\prime}_3}{2}\right)\sin^2\varphi\,,\\
\nonumber\sin^2\theta_{12}&=&\frac{\cos^2\varphi}{1-\left(\cos^2\rho\cos^2\theta+\sin^2\rho\sin^2\theta-\frac{1}{2}\sin2\rho\sin2\theta
\cos\frac{\kappa^{\prime}_2-\kappa^{\prime}_3}{2}\right)\sin^2\varphi}\,,\\
\nonumber\sin^2\theta_{23}&=&\Big\{\sin^2\phi\sin^2\rho+\cos^2\varphi\cos^2\phi\cos^2\rho+\frac{1}{2}\cos\varphi\sin2\phi\sin2\rho\cos2\theta\\
\nonumber&&-\sin^2\theta\cos2\rho
\left(\cos^2\phi\cos^2\varphi-\sin^2\phi\right)\\
\nonumber&&+\frac{1}{2}\cos\frac{\kappa^{\prime}_2-\kappa^{\prime}_3}{2}\left[\cos\varphi\sin2\phi\cos2\rho+\left(\sin^2\phi-\cos^2\phi\cos^2\varphi\right)\sin2\rho\right]\sin2\theta\Big\}\Big/\\
\nonumber&&\Big[1-\left(\cos^2\rho\cos^2\theta+\sin^2\rho\sin^2\theta-\frac{1}{2}\sin2\rho\sin2\theta\cos\frac{\kappa^{\prime}_2-\kappa^{\prime}_3}{2}\right)\sin^2\varphi\Big]\,,\\
\nonumber J_{CP}&=&\frac{1}{4}\cos\varphi\sin^2\varphi\sin2\phi\sin2\theta\sin\frac{\kappa^{\prime}_2-\kappa^{\prime}_3}{2}\,,\\
\nonumber |I_1|&=&\frac{1}{4}\sin^22\varphi\left|\cos^2\rho\sin^2\theta\sin\kappa^{\prime}_2+\sin^2\rho\cos^2\theta\sin\kappa^{\prime}_3+\frac{1}{2}\sin2\rho\sin2\theta
\sin\frac{\kappa^{\prime}_2+\kappa^{\prime}_3}{2}\right|\,,\\
|I_2|&=&\frac{1}{4}\sin^4\varphi\left|\sin^22\rho\cos2\theta\sin (\kappa^{\prime}_2-\kappa^{\prime}_3)+\sin4\rho\sin2\theta\sin\frac{\kappa^{\prime}_2-\kappa^{\prime}_3}{2}\right|\,.
\end{eqnarray}
It is easy to see that $\theta_{12}$ and $\theta_{13}$ are correlated as
\begin{equation}
\label{eq:theta12_13_2nd}\sin^2\theta_{12}\cos^2\theta_{13}=\cos^2\varphi\,,
\end{equation}
which leads to
\begin{equation}
0.303\pi\leq\varphi\leq0.329\pi\,.
\end{equation}
The allowed regions of $\varphi$ and $\phi$ by present experimental data is shown in Fig.~\ref{fig:range}. Moreover, a relation among $\delta_{CP}$ and three mixing angles is found:
\begin{equation}
\label{eq:cosd2}\cos\delta_{CP}=\frac{\cos2\theta_{23}(\cos^2\theta_{12}-\sin^2\theta_{12}\sin^2\theta_{13})-\cos2\phi(\cos^2\theta_{12}+\sin^2\theta_{12}\sin^2\theta_{13})}
{\sin2\theta_{12}\sin2\theta_{23}\sin\theta_{13}}\,,
\end{equation}
which allows us to constrain the values of $\cos\delta_{CP}$ versus $\phi$ for three mixing angles in the experimentally preferred $3\sigma$ ranges, as is plotted in Fig.~\ref{fig:phi-cosd}. In the same fashion as section~\ref{subsec:1st_column}, we have zero CP violation $\sin\delta_{CP}=0$ for $\kappa_2=\kappa_2$ or $\theta=0$, $\pi/2$, $\pi$, $3\pi/2$. A sufficient condition of maximal CP violation $\cos\delta_{CP}=0$ for any value of $\theta$ is still given by Eq.~\eqref{eq:maximal_1st}, and correspondingly the lepton mixing parameters are constrained by this symmetry to be of the form
\begin{eqnarray}
\nonumber&\sin^2\theta_{12}=\frac{\cos^2\varphi}{1-\cos^2\theta\sin^2\varphi},\qquad \sin^2\theta_{13}=\cos^2\theta\sin^2\varphi, \\
\label{eq:PMNS_2nd_maximal}&\sin^2\theta_{23}=\frac{1}{2},\qquad \cos\delta_{CP}=\sin\alpha_{31}=0,\qquad \tan\alpha_{21}=-\tan\kappa^{\prime}_2\,,
\end{eqnarray}
The predictions for the $(\beta\beta)_{0\nu}-$decay effective mass $|m_{ee}|$ is displayed in Fig.~\ref{fig:mee}(b). In the end, we consider the scenario where the residual $Z_2$ flavor symmetry generated by $G_{R}=X_{R1}X^{\ast}_{R2}=X_{R2}X^{\ast}_{R1}$ is a subgroup of a finite flavor symmetry. Then the possible forms of the column vector $(\cos\varphi, \sin\varphi\cos\phi, \sin\varphi\sin\phi)^{T}$ fixed by $G_{R}$ would be strongly constrained~\cite{Fonseca:2014koa}. In order to be compatible with experimental data~\cite{Gonzalez-Garcia:2014bfa}, this column can only be
\begin{equation}
\label{eq:2nd_column_finite_flasy}\left(\begin{array}{c}
\cos^2\varphi\\
\sin^2\varphi\cos^2\phi\\
\sin^2\varphi\sin^2\phi
\end{array}
\right)=\frac{1}{3}\left(\begin{array}{c}
1\\
1\\
1
\end{array}
\right)\quad\text{or}\quad\frac{1}{20}\left(\begin{array}{c}
10-2\sqrt{5} \\
5+\sqrt{5}  \\
5+\sqrt{5}
\end{array}
\right)\,.
\end{equation}
Both solutions lead to $\phi=\pi/4$. The predictions for $\sin^2\theta_{12}$ and $\cos\delta_{CP}$ by Eq.~\eqref{eq:theta12_13_2nd} and Eq.~\eqref{eq:cosd2} are listed in Table~\ref{tab:fixed_column}. We see that the solar mixing angle $\sin^2\theta_{12}$ is determined to be around 0.34 or 0.28 which can be checked at JUNO~\cite{JUNO}. Although $\cos\delta_{CP}$ is not constrained at $3\sigma$ level, it is found $-0.0382\leq\cos\delta_{CP}\leq0.734$ ($-1\leq\cos\delta_{CP}\leq-0.382$) or $-0.0439\leq\cos\delta_{CP}\leq0.843$ ($-1\leq\cos\delta_{CP}\leq-0.438$) for NO (IO) when both $\theta_{13}$ and $\theta_{23}$ lie in the $1\sigma$ regions.

\subsection{\label{subsec:3rd_column}Third column fixed for two remnant CP}

In this scenario, we can choose the permutation matrix is
\begin{equation}
P=\left(\begin{array}{ccc}
0    &  0    &   1   \\
1    &  0    &   0   \\
0    &  1    &   0
\end{array}
\right)\,.
\end{equation}
The PMNS matrix takes the form:
\begin{eqnarray}
\nonumber U_{PMNS}&=&\left(\begin{array}{ccc}
\sin\varphi    ~&~    0    ~&~   \cos\varphi      \\
-\cos\varphi\cos\phi    ~&~  \sin\phi   ~&~  \sin\varphi\cos\phi   \\
-\cos\varphi\sin\phi    ~&~  -\cos\phi  ~&~  \sin\varphi\sin\phi
\end{array}\right)
\left(\begin{array}{ccc}
\cos\rho   ~&~  \sin\rho   ~&~ 0  \\
-\sin\rho  ~&~  \cos\rho  ~&~0  \\
0   ~&~   ~0~  ~&~ 1
\end{array}\right)\\
&&\times\left(\begin{array}{ccc}
e^{i\frac{\kappa_2}{2}}   &  0   &   0    \\
0   &  e^{i \frac{\kappa_3}{2}}  &   0    \\
0   &   0  &   e^{i\frac{\kappa_1}{2}}
\end{array}\right)
\left(\begin{array}{ccc}
\cos\theta   ~&~  \sin\theta   ~&~  0  \\
-\sin\theta  ~&~  \cos\theta   ~&~  0  \\
0    ~&~   ~0~   ~&~  1
\end{array}
\right)\widehat{X}^{-1/2}_{R1}\,.
\end{eqnarray}
Now the vector $(\cos\varphi, \sin\varphi\cos\phi, \sin\varphi\sin\phi)^{T}$ resides in the third column of $U_{PMNS}$. The lepton mixing parameters read as
\begin{eqnarray}
\nonumber\sin^2\theta_{13}&=&\cos^2\varphi,\qquad \sin^2\theta_{23}=\cos^2\phi\,,\\
\nonumber\sin^2\theta_{12}&=&\frac{1}{2}\left(1-\cos2\rho\cos2\theta+\sin2\rho\sin2\theta\cos\frac{\kappa^{\prime}_2-\kappa^{\prime}_3}{2}\right)\,,\\
\nonumber\tan\delta_{CP}&=&\frac{\sin\frac{\kappa^{\prime}_2-\kappa^{\prime}_3}{2}}{\sin2\rho\cot2\theta+\cos2\rho\cos\frac{\kappa^{\prime}_2-\kappa^{\prime}_3}{2}}\,,\\
\nonumber\tan\alpha_{21}&=&-\frac{2\sin^22\rho\cos2\theta\sin(\kappa^{\prime}_2-\kappa^{\prime}_3)+2\sin4\rho\sin2\theta\sin\frac{\kappa^{\prime}_2-\kappa^{\prime}_3}{2}}
{(3\cos^22\rho-1)\sin^22\theta+\sin^22\rho(1+\cos^22\theta)\cos(\kappa^{\prime}_2-\kappa^{\prime}_3)+\sin4\rho\sin4\theta\cos\frac{\kappa^{\prime}_2-\kappa^{\prime}_3}{2}}\,,\\
\label{eq:mixing_parameters_3rd}\tan\alpha^{\prime}_{31}&=&-\frac{2\cos^2\rho\cos^2\theta\sin\kappa^{\prime}_2+2\sin^2\rho\sin^2\theta\sin\kappa^{\prime}_3-\sin2\rho\sin2\theta
\sin\frac{\kappa^{\prime}_2+\kappa^{\prime}_3}{2}}{2\cos^2\rho\cos^2\theta\cos\kappa^{\prime}_2+2\sin^2\rho\sin^2\theta\cos\kappa^{\prime}_3-\sin2\rho\sin2\theta
\cos\frac{\kappa^{\prime}_2+\kappa^{\prime}_3}{2}}\,.
\end{eqnarray}
The weak basis invariants are given by
{\small\begin{eqnarray}
\nonumber J_{CP}&=&\frac{1}{4}\cos\varphi\sin^2\varphi\sin2\phi\sin2\theta\sin\frac{\kappa^{\prime}_2-\kappa^{\prime}_3}{2}\,,\\
\nonumber
|I_1|&=&\frac{1}{4}\sin^4\varphi\left|\sin^22\rho\cos2\theta\sin(\kappa^{\prime}_2-\kappa^{\prime}_3)+\sin4\rho\sin2\theta\sin\frac{\kappa^{\prime}_2-\kappa^{\prime}_3}{2}\right|\,,\\
|I_2|&=&\frac{1}{4}\sin^22\varphi\left|2\cos^2\rho\cos^2\theta\sin\kappa^{\prime}_2+2\sin^2\rho\sin^2\theta\sin\kappa^{\prime}_3-\sin2\rho\sin2\theta
\sin\frac{\kappa^{\prime}_2+\kappa^{\prime}_3}{2}\right|\,.
\end{eqnarray}}
In this case, the allowed values of $\varphi$ and $\phi$ are strongly constrained by the measured values of $\theta_{13}$ and $\theta_{23}$:
\begin{equation}
0.449\pi\leq\varphi\leq0.456\pi,\qquad 0.204\pi\leq\phi\leq0.287\pi\,,
\end{equation}
at $3\sigma$ confidence level~\cite{Gonzalez-Garcia:2014bfa}. From the expression of $\tan\delta_{CP}$ in Eq.~\eqref{eq:mixing_parameters_3rd}, we see that Dirac CP would be maximally violated once the condition
\begin{equation}
\cot2\theta=-\cot2\rho\cos\frac{\kappa^{\prime}_2-\kappa^{\prime}_3}{2}\,,
\end{equation}
is satisfied. In particular, the parameters
\begin{equation}
\label{eq:maximal_3rd}\kappa_3=\kappa_2\pm\pi,\qquad
\rho=0,\frac{\pi}{2}, \pi~\text{or}~\frac{3\pi}{2}\,,
\end{equation}
lead to a maximal CP-violating phase with $\cos\delta_{CP}=0$ for any value of $\theta$. The associated remnant CP transformations take the form
{\small
\begin{eqnarray}
\nonumber X_{R1}&=&e^{i\kappa_+}\left(
\begin{array}{ccc}
 e^{i\frac{\kappa_2^{\prime}}{2}}-2ic^2_\varphi\sin\frac{\kappa_2^{\prime}}{2} &~ -i\sin2\varphi\cos\phi\sin\frac{\kappa_2^{\prime}}{2} ~& -i\sin2\varphi\sin\phi\sin\frac{\kappa_2^{\prime}}{2} \\
 -i\sin2\varphi\cos\phi\sin\frac{\kappa_2^{\prime}}{2} &~ \cos2\phi\,e^{i\frac{\kappa_2^{\prime}}{2}}-2i s^2_{\varphi}c^2_{\phi}\sin\frac{\kappa_2^{\prime}}{2} ~& \sin2\phi\left(e^{i\frac{\kappa_2^{\prime}}{2}}-is^2_\varphi\sin\frac{\kappa_2^{\prime}}{2}\right)\\
 -i\sin2\varphi\sin\phi\sin\frac{\kappa_2^{\prime}}{2} &~ \sin2\phi\left(e^{i\frac{\kappa_2^{\prime}}{2}}-is^2_\varphi\sin\frac{\kappa_2^{\prime}}{2}\right) ~&-\cos2\phi\,e^{i\frac{\kappa_2^{\prime}}{2}}-2is^2_{\varphi}s^2_{\phi}\sin\frac{\kappa_2^{\prime}}{2}\\
\end{array}
\right)\,,\\
\nonumber X_{R2}&=&e^{i\kappa_+}\left(
\begin{array}{ccc}
 e^{-i\frac{\kappa_2^{\prime}}{2}}-2s^2_\varphi\cos\frac{\kappa_2^{\prime}}{2} &~ \sin2\varphi\cos\phi\cos\frac{\kappa_2^{\prime}}{2} ~& \sin2\varphi\sin\phi\cos\frac{\kappa_2^{\prime}}{2} \\
 \sin2\varphi\cos\phi\cos\frac{\kappa_2^{\prime}}{2} & ~ -\cos2\phi\,e^{i\frac{\kappa_2^{\prime}}{2}}+2s^2_{\varphi}c^2_{\phi}\cos\frac{\kappa_2^{\prime}}{2}~ &-\sin2\phi\left(e^{-i\frac{\kappa_2^{\prime}}{2}}-s^2_\varphi\cos\frac{\kappa_2^\prime}{2}\right)\\
 \sin2\varphi\sin\phi\cos\frac{\kappa_2^{\prime}}{2} &~ -\sin2\phi\left(e^{-i\frac{\kappa_2^{\prime}}{2}}-s^2_\varphi\cos\frac{\kappa_2^\prime}{2}\right)~ &\cos2\phi\,e^{i\frac{\kappa_2^{\prime}}{2}}+2s^2_{\varphi}s^2_{\phi}\cos\frac{\kappa_2^{\prime}}{2} \\
\end{array}
\right)\,,
\end{eqnarray}}
for $\rho=0,~\pi$ and $\kappa_3=\kappa_2\pm\pi$,
where $c_{\varphi}$, $s_{\varphi}$, $c_{\phi}$ and $s_{\phi}$ are the abbreviations of $\cos\varphi$, $\sin\varphi$, $\cos\phi$ and $\sin\phi$ respectively. In case of $\rho=\pi/2,~3\pi/2$ and $\kappa_3=\kappa_2\pm\pi$, the above expressions for $X_{R1}$ and $X_{R2}$ are exchanged. Beside maximal $\delta_{CP}$, the Majorana phase $\alpha_{21}$ is enforced to be zero by the chosen residual CP while $\alpha_{31}$ is not constrained, i.e.,
\begin{equation}
\label{eq:PMNS_3rd_maximal}\cos\delta_{CP}=\sin\alpha_{21}=0,\qquad\tan\alpha_{31}=-\tan\kappa^{\prime}_2\,.
\end{equation}
The resulting predictions for $|m_{ee}|$ are displayed in Fig.~\ref{fig:mee}(c). Finally if the residual flavor symmetry $G_{R}=X_{R1}X^{\ast}_{R2}=X_{R2}X^{\ast}_{R1}$ originates from a finite flavor symmetry group, we find that only the third column of $\mathcal{C}_2$ is viable, i.e.,
\begin{equation}
\label{eq:3rd_column_finite_flasy}\left(\begin{array}{c}
\cos^2\varphi\\
\sin^2\varphi\cos^2\phi\\
\sin^2\varphi\sin^2\phi
\end{array}
\right)=\frac{2}{3}\left(
\begin{array}{c}
2\sin^2\theta_{\nu}  \\
2\sin^2\left(\theta_{\nu}-\frac{\pi}{3}\right) \\
2 \sin^2\left(\theta_{\nu}+\frac{\pi}{3}\right)
\end{array}
\right)\,,
\end{equation}
where $\theta_{\nu}\in\pm[0.0537\pi, 0.0622\pi]$ to be in accordance
with experimental data. In this case, $\theta_{13}$ and $\theta_{23}$ are determined to be: $0.0188\leq\sin^2\theta_{13}\leq0.0251$ and $0.387\leq\sin^2\theta_{23}\leq0.403$ or $0.597\leq\sin^2\theta_{23}\leq0.613$. The predictions for $\theta_{23}$ can be tested by forthcoming long baseline neutrino oscillation experiments.

\subsection{\label{subsec:single} Single remnant CP}

As demonstrated in section~\ref{sec:reconstruction_PMNS}, if the neutrino mass matrix is invariant under the action of a generic residual CP transformation $X_{R}=e^{i\kappa_1}v_1v^{T}_1+e^{i\kappa_2}v_2v^{T}_2+e^{i\kappa_3}v_3v^{T}_3$ in the charged lepton diagonal basis, then the lepton mixing matrix would be of the form
\begin{equation}
\label{eq:PMNS_one_CP_again}U_{PMNS}=\left(v_1, v_2, v_3\right)\text{diag}\big(e^{i\frac{\kappa_1}{2}}, e^{i\frac{\kappa_2}{2}}, e^{i\frac{\kappa_3}{2}}\big)O_{3\times3}\widehat{X}^{-\frac{1}{2}}_{R}\,.
\end{equation}
where $O_{3\times3}$ given by Eq.~\eqref{eq:Orthogonal_matrix} denotes an arbitrary orthogonal matrix. In this scenario, $U_{PMNS}$ depends on three free parameters $\theta_{1,2,3}$ besides the input parameters characterizing the residual CP transformation. The analytical expressions for the mixing parameters are rather lengthy and hence are omitted here. Since generally $\theta_{1,2,3}$ are involved in each entry of $U_{PMNS}$, the observed values of the three mixing angles can be easily accommodated by a suitable choice of the values of $\theta_{1,2,3}$. Depending on the concrete form of the residual CP transformation and $\theta_{1,2,3}$, the CP violating phases $\delta_{CP}$, $\alpha_{21}$ and $\alpha_{31}$ can take any values. After some cumbersome algebraic calculations, we find that $\delta_{CP}$ would be maximal for any $\theta_{1,2,3}$ if and only if
\begin{equation}
\label{eq:maximal_1CP}X_{R}=\left(
\begin{array}{ccc}
e^{i\kappa_a} & 0 & 0 \\
 0 & 0 & e^{i\kappa_b} \\
 0 & e^{i\kappa_b} & 0
\end{array}
\right)\,,
\end{equation}
where $a,b=1, 2, 3$. This is a minor generalization of the $\mu-\tau$ reflection symmetry in which $\kappa_a=\kappa_b$. Moreover, $\theta_{23}$ is enforced to be maximal by this residual CP transformation, and both Majorana phases are trivial, i.e.,
\begin{equation}
\label{eq:maximal_single}\sin^2\theta_{23}=\frac{1}{2},\qquad \cos\delta_{CP}=\sin\alpha_{21}=\sin\alpha_{31}=0\,,
\end{equation}
while $\theta_{12}$ and $\theta_{13}$ are not constrained. Similarly $\delta_{CP}$ will be zero (or equal to $\pi$) for any $\theta_{1,2,3}$ if
\begin{equation}
\label{eq:minimal_1CP}X_{R}=\left(\begin{array}{ccc}
e^{i\kappa_j}    &         0           &        0          \\
0                &   e^{i\kappa_m}     &        0          \\
0                &         0           &   e^{i\kappa_n}
\end{array}
\right)\,,
\end{equation}
where $j, m, n=1, 2, 3$. The Majorana CP-violating phases are also found to be conserved in this case,
\begin{equation}
\label{eq:minimal_single}\sin\delta_{CP}=\sin\alpha_{21}=\sin\alpha_{31}=0\,.
\end{equation}
If the residual CP is distinct from Eq.~\eqref{eq:maximal_1CP} and Eq.~\eqref{eq:minimal_1CP}, the Dirac CP would be neither conserved nor maximally violated except for some special values of $\theta_{1,2,3}$. From the formula of $|m_{ee}|$ in Eq.~\eqref{eq:mee}, we see that the two patterns in Eq.~\eqref{eq:maximal_single}(for conserved $\delta_{CP}$) and Eq.~\eqref{eq:minimal_single} (for maximal $\delta_{CP}$) leads to the same predictions for the effective mass $|m_{ee}|$, as is shown in Fig.~\ref{fig:mee}(d).

\section{\label{sec:conclusion}Summary and conclusions}

Over the past years, much effort has been devoted to understanding lepton flavor mixing angles from some discrete flavor symmetry. In this setup, the mismatch between the residual flavor symmetries in the neutrino and the charged lepton sectors generates the lepton mixing matrix. To account for the observed sizable $\theta_{13}$, the order of the flavor symmetry group should be quite large, the Dirac CP violating phase $\delta_{CP}$ is predicted to be conserved while the Majorana phases can not be determined by flavor symmetry alone~\cite{Fonseca:2014koa}. In this work, we propose to constrain the lepton flavor mixing matrix from residual CP symmetry instead of residual flavor symmetry.

The remnant CP symmetry can be derived from the experimentally measured mixing matrix. In the charged lepton diagonal basis, the neutrino mass matrix is invariant under the action of CP transformation of the neutrino triplets: $\nu_{L}(x)\stackrel{CP}{\longmapsto}iX_{\nu}\gamma^{0}C\bar{\nu}^{T}_{L}(x_P)$ and $X_{\nu}=U_{PMNS}\,\text{diag}(\pm1, \pm1, \pm1)U^{T}_{PMNS}$, and the remnant CP transformation of the charged lepton fields is a generic diagonal phase matrix. Performing two remnant CP transformations in succession can generate the well-known residual flavor Klein group $U_{PMNS}\,\text{diag}(\pm1, \pm1, \pm1)U^{\dagger}_{PMNS}$. As a result, the residual CP is more efficient than the residual flavor symmetry in predicting the lepton flavor mixing.

On the other hand, we have showed that the lepton mixing matrix can be constructed from the postulated residual CP transformations of the neutrino mass matrix. If the whole set of the residual CP transformations or three of them are preserved, lepton mixing angles and Dirac phase are completely fixed by the induced residual flavor symmetry. In addition, the Majorana CP phases are also subject to the constraint of remnant CP. In the case that there are two remnant CP transformations $X_{R1}$ and $X_{R2}$ in the neutrino sector, a by-product is a residual $Z_2$ flavor symmetry $G_{R}\equiv X_{R1}X^{\ast}_{R2}=X_{R2}X^{\ast}_{R1}$. As a result, one column of the PMNS matrix would be fixed by $G_{R}$. The PMNS matrix is found to be of the form of Eq.~\eqref{eq:PMNS_2CP}. We see that $U_{PMNS}$ depends on a single real free parameter $\theta$ besides the input parameters $\varphi$, $\phi$, $\rho$ and $\kappa_{1,2,3}$ specifying $X_{R1}$ and $X_{R2}$. Furthermore, if only one remnant CP transformation is kept by the neutrino mass matrix, the PMNS matrix is reconstructed to be given by Eq.~\eqref{eq:PMNS_one_CP}. $U_{PMNS}$ is determined up to an orthogonal matrix and it involves three real free parameters $\theta_{1,2,3}$. All the three CP violating phases $\delta_{CP}$, $\alpha_{21}$ and $\alpha_{31}$ are in general related to both the free parameter $\theta$ (or $\theta_{1,2,3}$) and the the parameters characterizing the remnant CP transformations, depending on the values of these parameters, they can take any values from 0 to $2\pi$.

The lepton masses can not be predicted in this approach. Therefore the column vector $(\cos\varphi, \sin\varphi\cos\phi, \sin\varphi\sin\phi)^{T}$ determined by the induced flavor symmetry $G_{R}$ can be in the first column, the second column or the third column of $U_{PMNS}$ in case of two remnant CP transformations. The phenomenological predictions of the mixing parameters are studied for each arrangement. The allowed regions of $\varphi$ and $\phi$ by the experimental data are extracted. Interesting relations of Eqs.~(\ref{eq:cosd1},\ref{eq:cosd2}) among $\delta_{CP}$ and mixing angles are found. We see that refined measurements of mixing angles can help to narrow or eventually pin down $\delta_{CP}$. In view of the preliminary result of $\delta_{CP}\sim3\pi/2$ from T2K, we have searched for the conditions of maximal Dirac CP violation for any value of $\theta$ (or $\theta_{1,2,3}$). Note that meanwhile at least one of the phases $\alpha_{21}$, $\alpha_{31}$ and $\alpha_{21}-\alpha_{31}$ is predicted to be zero or $\pi$. The corresponding predictions for the $(\beta\beta)_{0\nu}-$decay effctive mass $|m_{ee}|$ are studied. Furthermore, the scenario of induced flavor symmetry $G_{R}$ arising from a finite flavor symmetry group is discussed. The possible form of the column dictated by $G_{R}$ would be strongly constrained in this case, and consequently the mixing angles are found to lie in quite narrow regions. Comparison with forthcoming experimental data should be able to test this scenario.

In this paper the phenomenological implications of residual CP transformations are analyzed in a model-independent way. The remnant CP transformations can be any well-defined ones. The physical results only depend on the presumed remnant CP and are independent of how the remnant symmetry is dynamically realized. The idea of combining a flavor symmetry
with CP symmetry has recently stimulated some interesting discussions~\cite{Feruglio:2012cw,Ding:2013hpa,Feruglio:2013hia,Luhn:2013lkn,
Li:2013jya,Li:2014eia,Ding:2013bpa,Chen:2009gf,Girardi:2013sza,Ding:2013nsa,Ding:2014ssa,King:2014rwa,Hagedorn:2014wha,Ding:2014ora}, the predictions of the lepton mixing matrix for different symmetry breaking chains can be straightforwardly extracted via Eq.~\eqref{eq:PMNS_2CP} and Eq.~\eqref{eq:PMNS_one_CP} reported here. Finally it is interesting to explore the physical consequence of CP symmetry in some extension of the standard model such as the grand unification theory, where the viable CP transformations at high energy scale as well as remnant CP are strongly constrained by the gauge symmetry. In addition, it is possible to relate the leptonic CP violating phases to the precisely measured quark CP phase.

\section*{Acknowledgements}

This work is supported by the National Natural Science Foundation of China under Grant Nos. 11275188 and 11179007.


\begin{thebibliography}{99}

\bibitem{Harrison:2002er}
  P.~F.~Harrison, D.~H.~Perkins and W.~G.~Scott,
  %``Tri-bimaximal mixing and the neutrino oscillation data,''
 Phys.\ Lett.\ B {\bf 530}, 167 (2002)  [hep-ph/0202074];
%\bibitem{Harrison:2002kp}
  P.~F.~Harrison and W.~G.~Scott,
  %``Symmetries and generalizations of tri - bimaximal neutrino mixing,''
Phys.\ Lett.\ B {\bf 535}, 163 (2002)  [hep-ph/0203209];
%\bibitem{Xing:2002sw}
  Z.~z.~Xing,
  %``Nearly tri bimaximal neutrino mixing and CP violation,''
Phys.\ Lett.\ B {\bf 533} (2002) 85  [hep-ph/0204049];
%\bibitem{He:2003rm}
  X.~G.~He and A.~Zee,
  %``Some simple mixing and mass matrices for neutrinos,''
Phys.\ Lett.\ B {\bf 560}, 87 (2003)  [hep-ph/0301092].



\bibitem{Altarelli:2010gt}
  G.~Altarelli and F.~Feruglio,
  %``Discrete Flavor Symmetries and Models of Neutrino Mixing,''
  Rev.\ Mod.\ Phys.\  {\bf 82}, 2701 (2010)
  [arXiv:1002.0211 [hep-ph]].


\bibitem{Ishimori:2010zr}
  H.~Ishimori, T.~Kobayashi, H.~Ohki, Y.~Shimizu, H.~Okada and M.~Tanimoto,
  %``Non-Abelian Discrete Symmetries in Particle Physics,''
  Prog.\ Theor.\ Phys.\ Suppl.\  {\bf 183}, 1 (2010)
  [arXiv:1003.3552 [hep-th]].


\bibitem{Grimus:2012dk}
  W.~Grimus and P.~O.~Ludl,
  %``Finite family groups of fermions,''
  J.\ Phys.\ A {\bf 45}, 233001 (2012)
  [arXiv:1110.6376 [hep-ph]].


\bibitem{King:2013eh}
  S.~F.~King and C.~Luhn,
  %``Neutrino Mass and Mixing with Discrete Symmetry,''
  Rept.\ Prog.\ Phys.\  {\bf 76} (2013) 056201
  [arXiv:1301.1340 [hep-ph]].
  %%CITATION = ARXIV:1301.1340;%%


\bibitem{King:2014nza}
  S.~F.~King, A.~Merle, S.~Morisi, Y.~Shimizu and M.~Tanimoto,
  %``Neutrino Mass and Mixing: from Theory to Experiment,''
  arXiv:1402.4271 [hep-ph].
  %%CITATION = ARXIV:1402.4271;%%


\bibitem{Fonseca:2014koa}
  R.~M.~Fonseca and W.~Grimus,
  %``Classification of lepton mixing matrices from finite residual symmetries,''
  JHEP {\bf 1409}, 033 (2014)
  [arXiv:1405.3678 [hep-ph]].
  %%CITATION = ARXIV:1405.3678;%%


\bibitem{Abe:2011sj}
  K.~Abe {\it et al.}  [T2K Collaboration],
  %``Indication of Electron Neutrino Appearance from an Accelerator-produced Off-axis Muon Neutrino Beam,''
  Phys.\ Rev.\ Lett.\  {\bf 107}, 041801 (2011)
  [arXiv:1106.2822 [hep-ex]];
%\bibitem{Abe:2013fuq}
% K.~Abe {\it et al.}  [T2K Collaboration],
  %``Measurement of Neutrino Oscillation Parameters from Muon Neutrino Disappearance with an Off-axis Beam,''
  Phys.\ Rev.\ Lett.\  {\bf 111}, no. 21, 211803 (2013)
  [arXiv:1308.0465 [hep-ex]];
%\bibitem{Abe:2013hdq}
%  K.~Abe {\it et al.}  [T2K Collaboration],
  %``Observation of Electron Neutrino Appearance in a Muon Neutrino Beam,''
  Phys.\ Rev.\ Lett.\  {\bf 112}, 061802 (2014)
  [arXiv:1311.4750 [hep-ex]].


\bibitem{Adamson:2011qu}
  P.~Adamson {\it et al.}  [MINOS Collaboration],
  %``Improved search for muon-neutrino to electron-neutrino oscillations in MINOS,''
  Phys.\ Rev.\ Lett.\  {\bf 107}, 181802 (2011)
  [arXiv:1108.0015 [hep-ex]];
%\bibitem{Adamson:2013ue}
%  P.~Adamson {\it et al.}  [MINOS Collaboration],
  %``Electron neutrino and antineutrino appearance in the full MINOS data sample,''
  Phys.\ Rev.\ Lett.\  {\bf 110}, 171801 (2013)
  [arXiv:1301.4581 [hep-ex]];
%\bibitem{Adamson:2013whj}
%  P.~Adamson {\it et al.}  [MINOS Collaboration],
  %``Measurement of Neutrino and Antineutrino Oscillations Using Beam and Atmospheric Data in MINOS,''
  Phys.\ Rev.\ Lett.\  {\bf 110}, 251801 (2013)
  [arXiv:1304.6335 [hep-ex]].


\bibitem{Abe:2011fz}
  Y.~Abe {\it et al.}  [DOUBLE-CHOOZ Collaboration],
  %``Indication for the disappearance of reactor electron antineutrinos in
%  the Double Chooz experiment,''
  Phys.\ Rev.\ Lett.\  {\bf 108}, 131801 (2012)  [arXiv:1112.6353 [hep-ex]];
  %%CITATION = ARXIV:1112.6353;%%
%\bibitem{Abe:2012tg}
%  Y.~Abe {\it et al.}  [Double Chooz Collaboration],
  %``Reactor electron antineutrino disappearance in the Double Chooz
%  experiment,''
  Phys.\ Rev.\ D {\bf 86}, 052008 (2012)  [arXiv:1207.6632 [hep-ex]];
%\bibitem{Abe:2014lus}
%  Y.~Abe {\it et al.}  [Double Chooz Collaboration],
  %``Background-independent measurement of $\theta_{13}$ in Double Chooz,''
  Physics Letters B, Volume 735, 30 July 2014, Pages 51-56
  [arXiv:1401.5981 [hep-ex]];
%\bibitem{Abe:2014bwa}
  Y.~Abe {\it et al.}  [Double Chooz Collaboration],
  %``Improved measurements of the neutrino mixing angle $\theta_{13}$ with the Double Chooz detector,''
  JHEP {\bf 1410}, 86 (2014)
  [arXiv:1406.7763 [hep-ex]].


\bibitem{Ahn:2012nd}
  J.~K.~Ahn {\it et al.}  [RENO Collaboration],
  %``Observation of Reactor Electron Antineutrino Disappearance in the RENO
%  Experiment,''
  Phys.\ Rev.\ Lett.\  {\bf 108}, 191802 (2012)  [arXiv:1204.0626 [hep-ex]].


\bibitem{An:2012eh}
  F.~P.~An {\it et al.}  [DAYA-BAY Collaboration],
  %``Observation of electron-antineutrino disappearance at Daya Bay,''
   Phys.\ Rev.\ Lett.\  {\bf 108}, 171803 (2012)  [arXiv:1203.1669
   [hep-ex]];  %%CITATION = ARXIV:1203.1669;%%
%\bibitem{An:2012bu}
%  F.~P.~An {\it et al.}  [Daya Bay Collaboration],
  %``Improved Measurement of Electron Antineutrino Disappearance at Daya
%  Bay,''
  Chin.\  Phys.\ C {\bf 37}, 011001 (2013)  [arXiv:1210.6327 [hep-ex]];
% \bibitem{An:2013zwz}
%  F.~P.~An {\it et al.}  [Daya Bay Collaboration],
  %``Spectral measurement of electron antineutrino oscillation amplitude and frequency at Daya Bay,''
  Phys.\ Rev.\ Lett.\  {\bf 112}, 061801 (2014)
  [arXiv:1310.6732 [hep-ex]].


\bibitem{Abe:2013hdq}
  K.~Abe {\it et al.}  [T2K Collaboration],
  %``Observation of Electron Neutrino Appearance in a Muon Neutrino Beam,''
  Phys.\ Rev.\ Lett.\  {\bf 112}, 061802 (2014)
  [arXiv:1311.4750 [hep-ex]].


\bibitem{Capozzi:2013csa}
  F.~Capozzi, G.~L.~Fogli, E.~Lisi, A.~Marrone, D.~Montanino and A.~Palazzo,
  %``Status of three-neutrino oscillation parameters, circa 2013,''
  arXiv:1312.2878 [hep-ph].


\bibitem{Forero:2014bxa}
  D.~V.~Forero, M.~Tortola and J.~W.~F.~Valle,
  %``Neutrino oscillations refitted,''
  arXiv:1405.7540 [hep-ph].


\bibitem{Gonzalez-Garcia:2014bfa}
  M.~C.~Gonzalez-Garcia, M.~Maltoni and T.~Schwetz,
  %``Updated fit to three neutrino mixing: status of leptonic CP violation,''
  JHEP {\bf 1411}, 052 (2014)
  [arXiv:1409.5439 [hep-ph]].


\bibitem{Adams:2013qkq}
  C.~Adams {\it et al.}  [LBNE Collaboration],
  %``Scientific Opportunities with the Long-Baseline Neutrino Experiment,''
  arXiv:1307.7335 [hep-ex];
  %\bibitem{Bass:2013vcg}
  M.~Bass {\it et al.}  [LBNE Collaboration],
  %``Baseline optimization for the measurement of CP violation and mass hierarchy in a long-baseline neutrino oscillation experiment,''
  arXiv:1311.0212 [hep-ex].


%\bibitem{LBNF}
%The LBNF~Project, %\url{https://web.fnal.gov/project/LBNF/SitePages/Home.aspx}


\bibitem{::2013kaa}
  S.~K.~Agarwalla {\it et al.}  [LAGUNA-LBNO Collaboration],
  %``The mass-hierarchy and CP-violation discovery reach of the LBNO long-baseline neutrino experiment,''
  JHEP {\bf 1405}, 094 (2014)
  [arXiv:1312.6520 [hep-ph]];
%  \bibitem{Agostino:2014qoa}
  L.~Agostino, B.~Andrieu, R.~Asfandiyarov, D.~Autiero, O.~Bésida, F.~Bay, R.~Bayes and A.~M.~Blebea-Apostu {\it et al.},
  %``LBNO-DEMO: Large-scale neutrino detector demonstrators for phased performance assessment in view of a long-baseline oscillation experiment,''
  arXiv:1409.4405 [physics.ins-det];
%  \bibitem{Agarwalla:2014tca}
  S.~K.~Agarwalla {\it et al.}  [LAGUNA-LBNO Collaboration],
  %``Optimised sensitivity to leptonic CP violation from spectral information: the LBNO case at 2300 km baseline,''
  arXiv:1412.0593 [hep-ph];  %%CITATION = ARXIV:1412.0593;%%
%\bibitem{Agarwalla:2014ura}
  S.~K.~Agarwalla {\it et al.}  [LAGUNA-LBNO Collaboration],
  %``The LBNO long-baseline oscillation sensitivities with two conventional neutrino beams at different baselines,''
  arXiv:1412.0804 [hep-ph].


\bibitem{Abe:2011ts}
  K.~Abe, T.~Abe, H.~Aihara, Y.~Fukuda, Y.~Hayato, K.~Huang, A.~K.~Ichikawa and M.~Ikeda {\it et al.},
  %``Letter of Intent: The Hyper-Kamiokande Experiment --- Detector Design and Physics Potential ---,''
  arXiv:1109.3262 [hep-ex];
%  \bibitem{Kearns:2013lea}
  E.~Kearns {\it et al.}  [Hyper-Kamiokande Working Group Collaboration],
  %``Hyper-Kamiokande Physics Opportunities,''
  arXiv:1309.0184 [hep-ex].


\bibitem{Ecker:1981wv}
  G.~Ecker, W.~Grimus and W.~Konetschny,
  %``Quark Mass Matrices In Left-right Symmetric Gauge Theories,''
  Nucl.\ Phys.\ B {\bf 191} (1981) 465;
%\bibitem{Ecker:1983hz}
  G.~Ecker, W.~Grimus and H.~Neufeld,
  %``Spontaneous Cp Violation In Left-right Symmetric Gauge Theories,''
  Nucl.\ Phys.\ B {\bf 247} (1984) 70;
%\bibitem{Ecker:1987qp}
  G.~Ecker, W.~Grimus and H.~Neufeld,
  %``A Standard Form For generalized CP Transformations,''
  J.\ Phys.\ A {\bf 20} (1987) L807;
%\bibitem{Neufeld:1987wa}
  H.~Neufeld, W.~Grimus and G.~Ecker,
  %``Generalized Cp Invariance, Neutral Flavor Conservation And The
%  Structure Of The Mixing Matrix,''
  Int.\ J.\ Mod.\ Phys.\ A {\bf 3}, 603 (1988).


\bibitem{Grimus:1995zi}
  W.~Grimus and M.~N.~Rebelo,
  %``Automorphisms in gauge theories and the definition of CP and P,''
  Phys.\ Rept.\  {\bf 281}, 239 (1997)
  [arXiv:9506272[hep-ph]].


\bibitem{Harrison:2002kp}
  P.~F.~Harrison and W.~G.~Scott,
  %``Symmetries and generalizations of tri - bimaximal neutrino mixing,''
  Phys.\ Lett.\ B {\bf 535}, 163 (2002)
  [hep-ph/0203209];
%\bibitem{Harrison:2002et}
  P.~F.~Harrison and W.~G.~Scott,
  %``mu - tau reflection symmetry in lepton mixing and neutrino
%  oscillations,''
  Phys.\ Lett.\ B {\bf 547}, 219 (2002)
  [hep-ph/0210197];
  %%CITATION = HEP-PH/0210197;%%
%\bibitem{Harrison:2004he}
  P.~F.~Harrison and W.~G.~Scott,
  %``The Simplest neutrino mass matrix,''
  Phys.\ Lett.\ B {\bf 594}, 324 (2004)
  [hep-ph/0403278].


\bibitem{Grimus:2003yn}
  W.~Grimus and L.~Lavoura,
  %``A Nonstandard CP transformation leading to maximal atmospheric neutrino
 % mixing,''
  Phys.\ Lett.\ B {\bf 579}, 113 (2004)
  [hep-ph/0305309];
%\bibitem{Grimus:2012hu}
  W.~Grimus and L.~Lavoura,
  %``mu-tau Interchange symmetry and lepton mixing,''
  Fortsch.\ Phys.\  {\bf 61}, 535 (2013)
  [arXiv:1207.1678 [hep-ph]].


\bibitem{Farzan:2006vj}
  Y.~Farzan and A.~Y.~.Smirnov,
  %``Leptonic CP violation: Zero, maximal or between the two extremes,''
  JHEP {\bf 0701}, 059 (2007)
  [hep-ph/0610337].


\bibitem{Mohapatra:2012tb}
  R.~N.~Mohapatra and C.~C.~Nishi,
  %``$S_4$ Flavored CP Symmetry for Neutrinos,''
  Phys.\ Rev.\ D {\bf 86}, 073007 (2012)
  [arXiv:1208.2875 [hep-ph]].


\bibitem{Feruglio:2012cw}
  F.~Feruglio, C.~Hagedorn and R.~Ziegler,
  %``Lepton Mixing Parameters from Discrete and CP Symmetries,''
  JHEP {\bf 1307}, 027 (2013)
  [arXiv:1211.5560 [hep-ph]].


\bibitem{Ding:2013hpa}
  G.~-J.~Ding, S.~F.~King, C.~Luhn and A.~J.~Stuart,
  %``Spontaneous CP violation from vacuum alignment in $S_4$ models of
%  leptons,''
  JHEP {\bf 1305}, 084 (2013)
  [arXiv:1303.6180 [hep-ph]].


\bibitem{Feruglio:2013hia}
  F.~Feruglio, C.~Hagedorn and R.~Ziegler,
  %``A realistic pattern of lepton mixing and masses from $S_4$ and CP,''
  Eur.\ Phys.\ J.\ C {\bf 74}, 2753 (2014)
  [arXiv:1303.7178 [hep-ph]].


\bibitem{Luhn:2013lkn}
  C.~Luhn,
  %``Trimaximal TM$_{1}$ neutrino mixing in S$_{4}$ with spontaneous CP violation,''
  Nucl.\ Phys.\ B {\bf 875}, 80 (2013)
  [arXiv:1306.2358 [hep-ph]].


 \bibitem{Li:2013jya}
  C.~-C.~Li and G.~-J.~Ding,
  %``generalized CP and trimaximal $TM_1$ lepton mixing in $S_4$ family symmetry,''
  Nucl.\ Phys.\ B {\bf 881}, 206 (2014)
  [arXiv:1312.4401 [hep-ph]].


\bibitem{Li:2014eia}
  C.~C.~Li and G.~J.~Ding,
  %``Deviation from Bimaximal Mixing and Leptonic CP Phases in $S_4$ Family Symmetry and Generalized CP,''
  arXiv:1408.0785 [hep-ph].


\bibitem{Ding:2013bpa}
  G.~-J.~Ding, S.~F.~King and A.~J.~Stuart,
  %``generalized CP and $A_4$ Family Symmetry,''
  JHEP {\bf 1312} (2013) 006
  [arXiv:1307.4212].


\bibitem{Chen:2009gf}
  M.~-C.~Chen and K.~T.~Mahanthappa,
  %``Group Theoretical Origin of CP Violation,''
  Phys.\ Lett.\ B {\bf 681}, 444 (2009)  [arXiv:0904.1721 [hep-ph]].


\bibitem{Girardi:2013sza}
  I.~Girardi, A.~Meroni, S.~T.~Petcov and M.~Spinrath,
  %``Generalised geometrical CP violation in a T' lepton flavour model,''
  JHEP {\bf 1402}, 050 (2014)
  [arXiv:1312.1966 [hep-ph]].


\bibitem{Ding:2013nsa}
  G.~-J.~Ding and Y.~-L.~Zhou,
  %``Predicting Lepton Flavor Mixing from $\Delta(48)$ and Generalized CP Symmetries,''
  arXiv:1312.5222 [hep-ph];
%\bibitem{Ding:2014hva}
  G.~-J.~Ding and Y.~-L.~Zhou,
  %``Lepton mixing parameters from $\Delta(48)$ family symmetry and generalized CP,''
  JHEP {\bf 1406}, 023 (2014)
  [arXiv:1404.0592 [hep-ph]].


\bibitem{Ding:2014ssa}
  G.~-J.~Ding and S.~F.~King,
  %``generalized CP and $\Delta (96)$ Family Symmetry,''
  Phys.\ Rev.\ D {\bf 89}, 093020 (2014)
  [arXiv:1403.5846 [hep-ph]].


\bibitem{King:2014rwa}
  S.~F.~King and T.~Neder,
  %``Lepton Mixing Predictions including Majorana Phases from $\Delta(6n^2)$ Flavour Symmetry and generalized CP,''
  arXiv:1403.1758 [hep-ph].


\bibitem{Hagedorn:2014wha}
 C.~Hagedorn, A.~Meroni and E.~Molinaro,
%``Lepton Mixing from $\Delta$ (3 $n^2$) and $\Delta$ (6 $n^2$) and CP,''
arXiv:1408.7118 [hep-ph].


\bibitem{Ding:2014ora}
  G.~J.~Ding, S.~F.~King and T.~Neder,
  %``Generalised CP and $\Delta(6n^2)$ family symmetry in semi-direct models of leptons,''
  JHEP {\bf 1412}, 007 (2014)
  [arXiv:1409.8005 [hep-ph]].


\bibitem{Holthausen:2012dk}
  M.~Holthausen, M.~Lindner and M.~A.~Schmidt,
  %``CP and Discrete Flavour Symmetries,''
  JHEP {\bf 1304}, 122 (2013)
  [arXiv:1211.6953 [hep-ph]].


%\cite{Chen:2014tpa}
\bibitem{Chen:2014tpa}
  M.~C.~Chen, M.~Fallbacher, K.~T.~Mahanthappa, M.~Ratz and A.~Trautner,
  %``CP Violation from Finite Groups,''
  Nucl.\ Phys.\ B {\bf 883}, 267 (2014)
  [arXiv:1402.0507 [hep-ph]].
  %%CITATION = ARXIV:1402.0507;%%


\bibitem{Branco:1983tn}
  G.~C.~Branco, J.~M.~Gerard and W.~Grimus,
  %``Geometrical T Violation,''
  Phys.\ Lett.\ B {\bf 136}, 383 (1984);  %%CITATION = PHLTA,B136,383;%%
%\bibitem{deMedeirosVarzielas:2011zw}
  I.~de Medeiros Varzielas and D.~Emmanuel-Costa,
  %``Geometrical CP Violation,''
  Phys.\ Rev.\ D {\bf 84}, 117901 (2011)  [arXiv:1106.5477 [hep-ph]]; %
% \bibitem{Varzielas:2012nn}
  I.~de Medeiros Varzielas, D.~Emmanuel-Costa and P.~Leser,
  %``Geometrical CP Violation from Non-Renormalisable Scalar Potentials,''
  Phys.\ Lett.\ B {\bf 716}, 193 (2012)  [arXiv:1204.3633 [hep-ph]];
%\bibitem{Varzielas:2012pd}
  I.~de Medeiros Varzielas,
  %``Geometrical CP violation in multi-Higgs models,''
  JHEP {\bf 1208}, 055 (2012)  [arXiv:1205.3780 [hep-ph]];
%\bibitem{Bhattacharyya:2012pi}
  G.~Bhattacharyya, I.~de Medeiros Varzielas and P.~Leser,
  %``A common origin of fermion mixing and geometrical CP violation, and its
%  test through Higgs physics at the LHC,''
  Phys.\ Rev.\ Lett.\  {\bf 109}, 241603 (2012)  [arXiv:1210.0545 [hep-ph]].


\bibitem{Branco:2011zb}
  G.~C.~Branco, R.~G.~Felipe and F.~R.~Joaquim,
  %``Leptonic CP Violation,''
  Rev.\ Mod.\ Phys.\  {\bf 84}, 515 (2012)  [arXiv:1111.5332 [hep-ph]].


\bibitem{Ge:2011ih}
  S.~F.~Ge, D.~A.~Dicus and W.~W.~Repko,
  %``Z_2 Symmetry Prediction for the Leptonic Dirac CP Phase,''
  Phys.\ Lett.\ B {\bf 702}, 220 (2011)  [arXiv:1104.0602 [hep-ph]];
%??????????????????????????????????
%\bibitem{Ge:2011qn}
  S.~F.~Ge, D.~A.~Dicus and W.~W.~Repko,
  %``Residual Symmetries for Neutrino Mixing with a Large $\theta_{13}$ and Nearly Maximal $\delta_D$,''
  Phys.\ Rev.\ Lett.\  {\bf 108}, 041801 (2012)  [arXiv:1108.0964 [hep-ph]];
%\bibitem{Hernandez:2012ra}
  D.~Hernandez and A.~Y.~Smirnov,
  %``Lepton mixing and discrete symmetries,''
  Phys.\ Rev.\ D {\bf 86}, 053014 (2012)  [arXiv:1204.0445 [hep-ph]];
%\bibitem{Hernandez:2012sk}
  D.~Hernandez and A.~Y.~Smirnov,
  %``Discrete symmetries and model-independent patterns of lepton mixing,''
  Phys.\ Rev.\ D {\bf 87}, no. 5, 053005 (2013)  [arXiv:1212.2149 [hep-ph]].


\bibitem{Jarlskog:1985ht}
  C.~Jarlskog,
  %``Commutator of the Quark Mass Matrices in the Standard Electroweak Model and a Measure of Maximal CP Violation,''
  Phys.\ Rev.\ Lett.\  {\bf 55}, 1039 (1985).


\bibitem{pdg}
  K.~A.~Olive {\it et al.}  [Particle Data Group Collaboration],
  %``Review of Particle Physics,''
  Chin.\ Phys.\ C {\bf 38}, 090001 (2014).
  %%CITATION = CHPHD,C38,090001;%%


\bibitem{Branco:1986gr}
  G.~C.~Branco, L.~Lavoura and M.~N.~Rebelo,
  %``Majorana Neutrinos and {CP} Violation in the Leptonic Sector,''
  Phys.\ Lett.\ B {\bf 180}, 264 (1986).
  %%CITATION = PHLTA,B180,264;%%


\bibitem{Jenkins:2007ip}
  E.~E.~Jenkins and A.~V.~Manohar,
  %``Rephasing Invariants of Quark and Lepton Mixing Matrices,''
  Nucl.\ Phys.\ B {\bf 792}, 187 (2008)
  [arXiv:0706.4313 [hep-ph]].


\bibitem{Auger:2012ar}
  M.~Auger {\it et al.}  [EXO Collaboration],
  %``Search for Neutrinoless Double-Beta Decay in $^{136}$Xe with EXO-200,''
  Phys.\ Rev.\ Lett.\  {\bf 109}, 032505 (2012)
  [arXiv:1205.5608 [hep-ex]].


\bibitem{Albert:2014awa}
  J.~B.~Albert {\it et al.}  [EXO-200 Collaboration],
  %``Search for Majorana neutrinos with the first two years of EXO-200 data,''
  Nature {\bf 510} (2014) 229-234
  [arXiv:1402.6956 [nucl-ex]].


\bibitem{Gando:2012zm}
  A.~Gando {\it et al.}  [KamLAND-Zen Collaboration],
  %``Limit on Neutrinoless $\beta\beta$ Decay of Xe-136 from the First
  %Phase of KamLAND-Zen and Comparison with the Positive Claim in
  %Ge-76,''
  Phys.\ Rev.\ Lett.\  {\bf 110} (2013) 062502
  [arXiv:1211.3863 [hep-ex]].


\bibitem{Ade:2013zuv}
  P.~A.~R.~Ade {\it et al.}  [Planck Collaboration],
  %``Planck 2013 results. XVI. Cosmological parameters,''
  Astron.\ Astrophys.\  {\bf 571}, A16 (2014)
  [arXiv:1303.5076 [astro-ph.CO]].


\bibitem{JUNO}JUNO experiment, \url{http://english.ihep.cas.cn/rs/fs/juno0815/}



\end{thebibliography}
\end{document}